\documentclass{aa}
\usepackage{graphicx}
\usepackage{txfonts}
\bibpunct{(}{)}{;}{a}{}{,} % to follow the A&A style
\begin{document}

   \title{Reconstruction of a helical prominence in 3D}

   \subtitle{from IRIS spectra and images}

   \author{B. Schmieder \inst{1}
          %\inst{1}
          \and
           M. Zapi\'or \inst{2}
                A. L\'opez Ariste \inst{3}
                \and  
          P. Levens \inst{4}
           \and 
           N. Labrosse  \inst{4}
                  \and 
          %\inst{1}
 %    (J.M. Malherbe \inst{1})
%         \and
          R. Gravet \inst{5}
%                 \and
  %\and
         %T.Roudier \inst{5}
          %\and
          %\inst{1}\fnmsep\thanks{Just to show the usage
          %of the elements in the author field}
          }
 \institute{LESIA, Observatoire de Paris, PSL Research University, CNRS, Sorbonne Universit\'{e}s, UPMC Univ. Paris 06, Univ. Paris-Diderot, Sorbonne Paris Cité, 5 place Jules Janssen, F-92195 Meudon, France\\
   \email{brigitte.schmieder@obspm.fr}
     \and
           Astronomical Institute, Academy of Sciences of the Czech Republic, Fri{\v c}ova 298, 25165 Ond{\v r}ejov, Czech Republic\\
           \email{zapior.maciej@gmail.com}
                 \and
                Institut de Recherche en Astrophysique et Plan\'{e}tologie, Toulouse, France\\
                \and
   SUPA School of Physics and Astronomy, University of Glasgow,
              Glasgow, G12 8QQ, UK\\
             % \email{p.levens.1@research.gla.ac.uk} \\
              \and
                Universit\'{e} de Orl\'{e}ans,  Orl\'{e}ans, France\\
                   %    University of Alexandria, Department of Geography, ...\\
         %    \email{c.ptolemy@hipparch.uheaven.space}
         %    \thanks{The university of heaven temporarily does not
         %            accept e-mails}
             }

   \date{Received ...; accepted ...}

% \abstract{}{}{}{}{} 
% 5 {} token are mandatory
 \abstract
   {Movies of prominences obtained  by space instruments e.g. the  Solar Optical Telescope (SOT) aboard the {\it Hinode} satellite and  the Interface Region Imaging Spectrograph (IRIS)  with high  temporal and spatial resolution revealed the tremendous dynamical nature of prominences. { Knots of plasma  belonging to prominences} appear to travel along both vertical  and horizontal  thread-like loops, with highly dynamical nature.}
   {The aim of the paper is to reconstruct the 3D shape of a helical prominence observed over two and a half hours  by IRIS.}
   {From  the IRIS \ion{Mg}{ii} k spectra we compute Doppler shifts of the plasma inside the prominence and from the slit-jaw images (SJI){ we  derive the transverse field in the plane of the sky. Finally  we obtain the velocity vector  field of the knots   in 3D.}}
   {{ We reconstruct the real trajectories of nine knots travelling along ellipses.}}
   {The  spiral-like structure of the prominence observed in the plane of the sky is mainly due to  the projection effect of  long arches of threads (up to 8 $ \times 10^4$ km). Knots run along more or less horizontal threads with velocities reaching 65 km s$^{-1}$. The dominant driving force is the gas pressure.}

   \keywords{Solar prominence,  dynamics, magnetic structure.}

   \maketitle

\section{Introduction}

%Figure 1

  \begin{figure*}
   \centering
\includegraphics[width=0.45\textwidth]{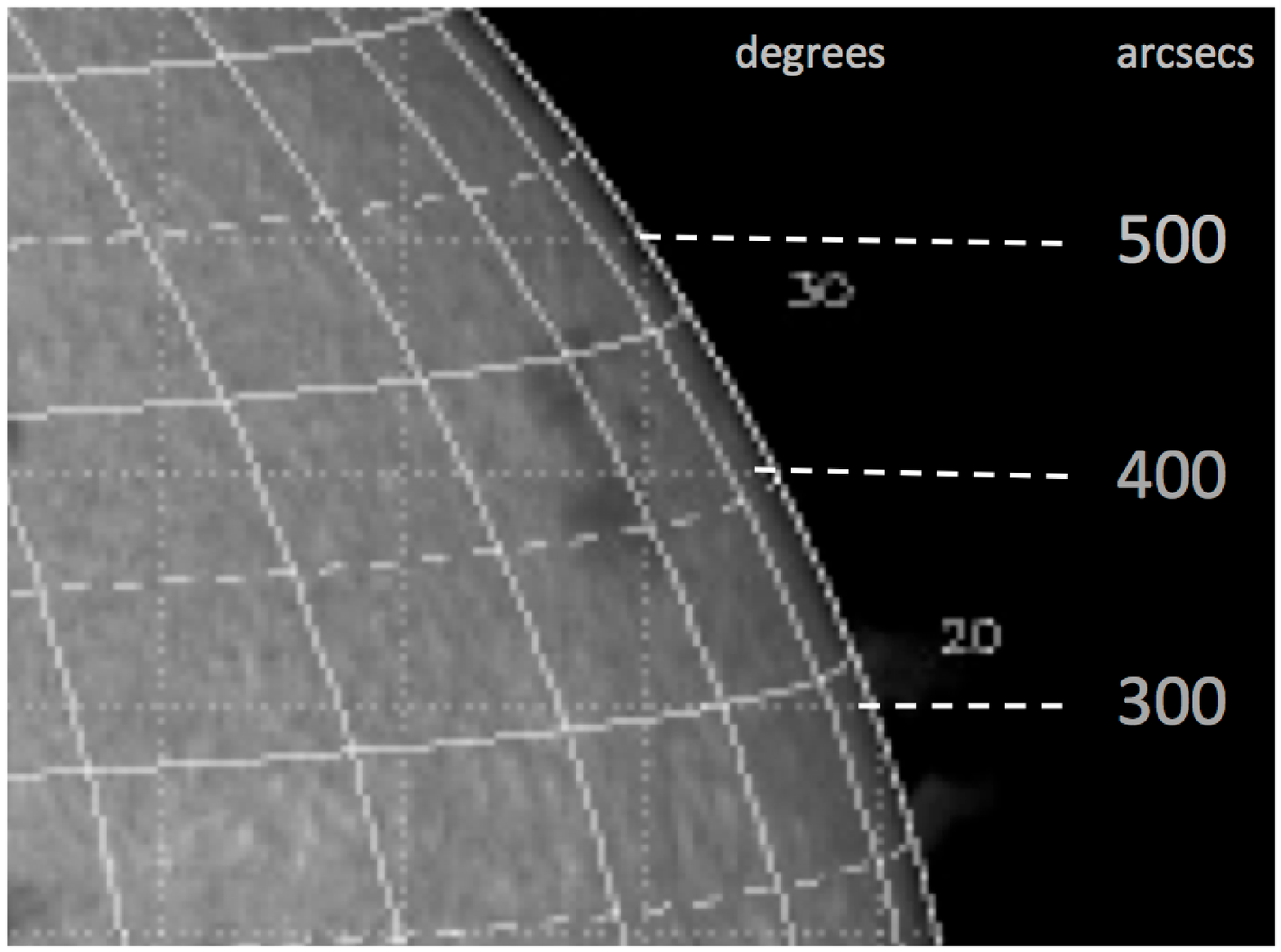}
   \includegraphics[width=0.45\textwidth]{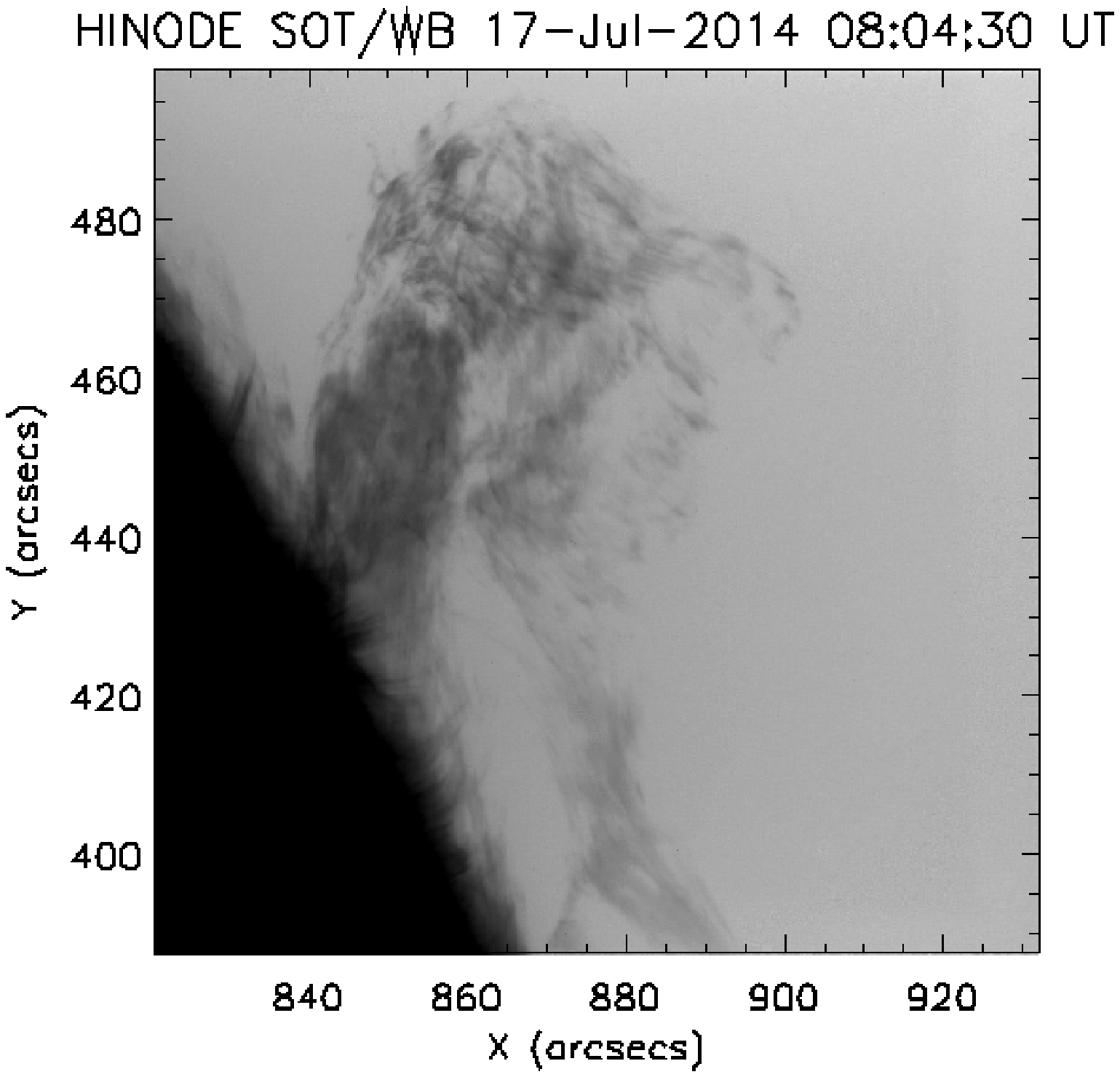}
      \caption{( Left panel:) Filament observed with the Meudon spectroheliograph in H$\alpha$ on July 15, 2014  at 07:20 UT.  ( Right panel:) The corresponding 
      filament (in reverse colour) as it crossed the limb as a prominence on July 17, 2014, observed by SOT using the \ion{Ca}{II}  filter  at 08:04 UT.  A movie of the  SOT image  is available online (Movie 1).} 
         \label{Meudon}
   \end{figure*}
%
  %Figure 2

  \begin{figure*}
   \centering
   \includegraphics[width=0.38\textwidth]{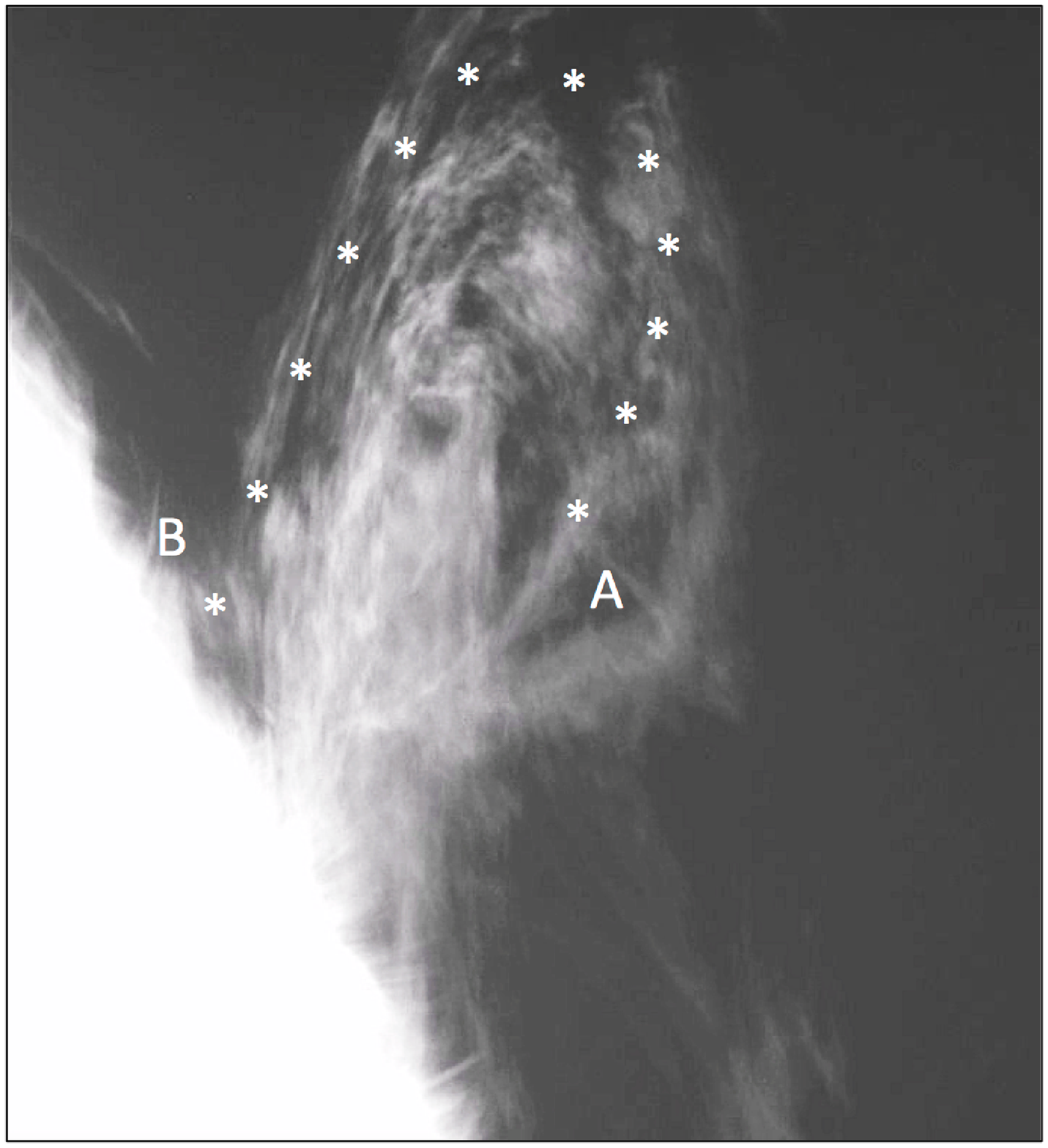}
    \includegraphics[width=0.48\textwidth]{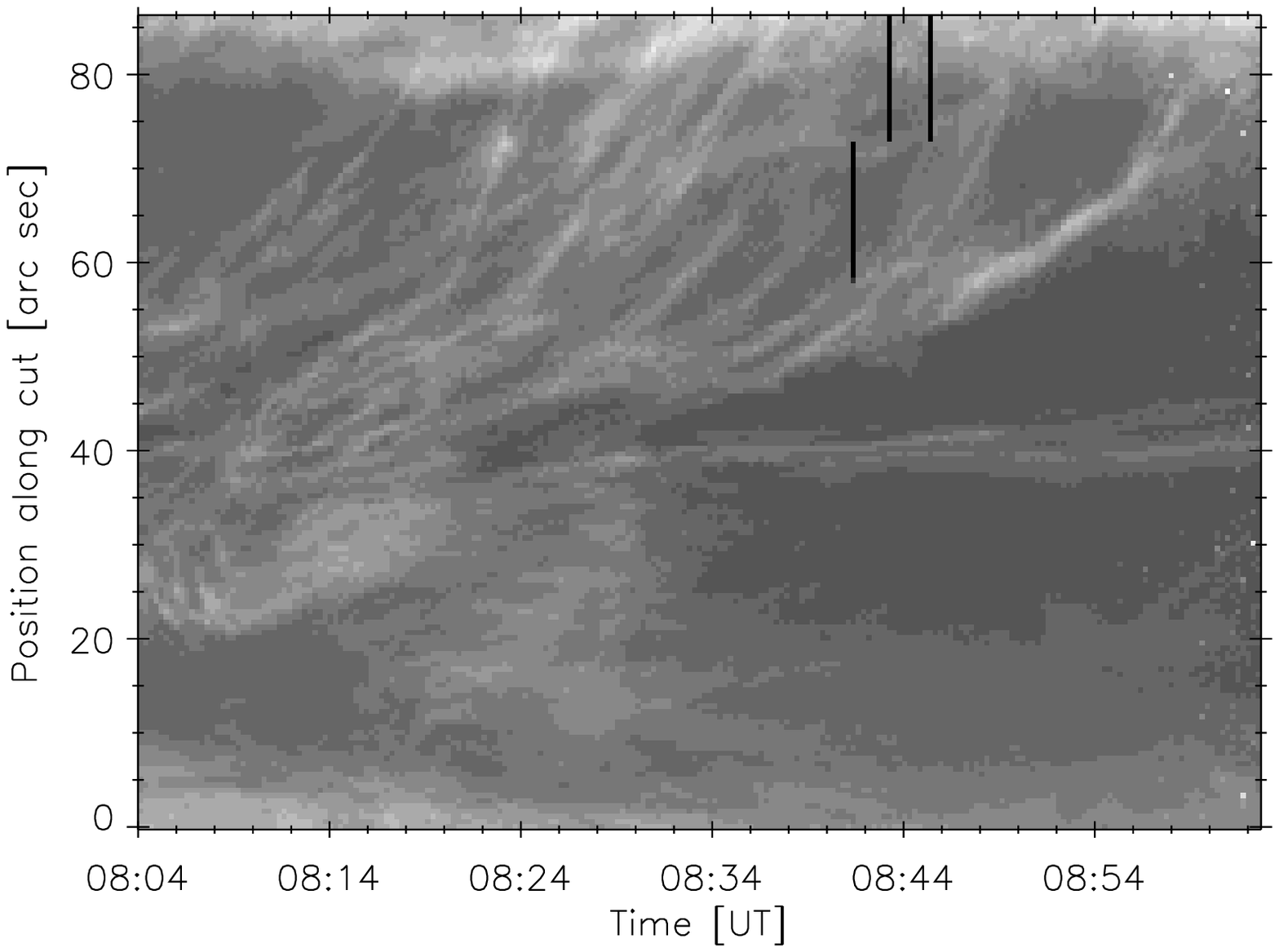}
      \caption{(Left panel:) Prominence observed by SOT  using the \ion{Ca}{II}  K filter at 08:10 UT on July 17. Crosses follow an  elliptical loop from right (A) to left (B). (Right panel:) Time-distance diagram for the path along the loop joining A and B. The black vertical lines in the diagram are missing data.
      }
       \label{time}
   \end{figure*}

The Solar Optical Telescope \citep[SOT;][]{tsuneta08,suematsu08}  aboard the \textit{Hinode} satellite \citep{kosugi07}  has observed prominences in H$\alpha$ and  the chromospheric \ion{Ca}{II} lines. It revealed the highly dynamical nature of cool plasma in prominences \citep{Labrosse10,Dudik2012}.
The \textit{Solar Dynamics Observatory} spacecraft and its high  spatial and temporal resolution imager, the Atmospheric Imaging Assembly \citep[\textit{SDO}, AIA;][]{Lemen12} 
allow us to follow the dynamics of prominences or filamenst when observed on the disk in transition-region and coronal filters  (304 \AA\  filter $\sim$ 10$^5$ K,   171 \AA\ and 193 \AA\  filters $\sim$ 10$^6$ K) \citep{Parenti12}. 

AIA movies reveal that some prominence structures  when  arriving close to the limb look like  tornadoes, with  an apparent rotation motion around their vertical axis 
\citep{Su12,Li12,panesar13,Su14,Levens2015}. 
%Another kind of tornado-like structure can be described by a large circular structure atop a narrow pillar with fine structures rotating in the top structure \citep{Li12}. 
 The Extreme-ultraviolet Imaging Spectrometer \citep[EIS;][]{Culhane07} on board \textit{Hinode} has provided spectral scans covering  prominence-tornado structures  in different wavelengths (170--211 \AA\ and 246--292 \AA).  The Dopplergrams of such structures presented  blueshifts on one side of the vertical structure and redshifts on the other side \citep{Su14,Levens2015}.  These former authors  used time-distance diagrams from coronal filtergrams taken with the  AIA to  quantify sinusoidal oscillations in tornadoes, finding      periods  of about one  hour and apparent rotational velocities of 6--8 $\mathrm{km s}^{-1}$.
A sit-and-stare EIS run  showed that this kind of pattern could survive for three hours suggesting  a long life of the rotation motion \citep{Su14}.
 
 These observations in hot lines  (> 10$^6$ K) concern  mainly the interface of cool prominences (10$^4$ K) with the corona.  It appears important to understand the dynamics of the cool plasma inside  these structures.  Attempts have been  made to derive the line-of-sight velocity of the cool plasma  using chromospheric lines observed  by ground-based telescopes.  Swirling motions in the low atmosphere were detected  using the Crisp Imaging Spectropolarimeter at the Swedish Solar Telescope \citep[CRISP, SST;][]{Scharmer03,Wedemeyer2014}.  However they  do not  appear to be related to filamentary structures. On the other hand,  high dark vertical  structures  observed at the limb with the CRISP,  which look  like  tornadoes  in AIA filters,  have been interpreted as    legs of prominences \citep{Wedemeyer13}.  Prominence tornadoes  have been observed in the \ion{He}{I} 10830 \AA\ line 
 with the  Tenerife Infrared Polarimeter at the Vacuum Tower Telescope  in Tenerife  \citep[TIP, VTT;][]{Orozco12,Martinez16}. Both sit-and-stare observations along  a slit  and scans of a region  with a  cadence of  half an  hour have been performed. Using the  sit and stare mode, \citet{Orozco12} found an    anti-symmetric Doppler  curve similar to the result found by  \citet{Su12} using EIS. 
  \citet{Martinez16} used the scanning  mode   and obtained   four consecutive spectro-polarimetric scans    in four hours.  The latter authors could not find any coherent behaviour between the two scans and concluded that, if rotation exists, it must be intermittent and last less than one hour.  In fact, the incoherence that they found has been explained 
 by observations of a tornado made  using the Meudon Solar Tower
 and  the Multi Subtractive Double Pass (MSDP)  spectrograph,  which provides  Dopplergrams  with a cadence of 30 seconds. Over a period of two hours  large cells of blueshifts and redshifts have been observed in a prominence \citep{Schmieder2017}. 
%However the blue cells were switching to red cells after 30 minutes and vice versa for the red cells. 
Over a period of 30 minutes it was found that redshifted cells became blueshifted, and vice versa.
They explained this quasi-periodicity in the Doppler shift maps  as being caused by oscillations of the  dipped magnetic structure sustaining the prominence plasma, and not by rotation of the structure.  In prominences  the magnetic field is parallel to the photosphere, with  cool plasma  suspended in dips \citep{Aulanier1998,Lopez2006,Dudik2008,Gunar2015,Gunar2016}. 
 Vector magnetograms have been obtained using the polarimeter at the T\'{e}lescope H\'{e}liographique pour l'Etude du Magn\'{e}tisme et des Instabilit\'{e}s Solaires (THEMIS, French telescope in the Canary Islands-see \citet{Rayrole2000,Bommier2005}), and they confirmed that the magnetic field  in tornado-like prominences is  roughly horizontal \citep[][ Levens et al. 2017 {\it in prep.}]{Schmieder2014,Levens2016a,Levens2016b}.
{  Helical motions have been  detected in  prominences \citep{Martinez2015,Zapior2016}.   For example,  knot trajectories along  elliptical loops have been computed using a method of 3D reconstruction  developed in \citet{Zapior2012}. We shall use the same method in the present    paper. } Such a kind of helical structure could correspond to the theoretical model of  a tornado  \citep{Luna2015}.
 
 During a  coordinated campaign   in July 2014, many prominences were observed using  the SOT instrument aboard \textit{Hinode},  and the Interface Region Imaging Spectrograph \citep[IRIS;][]{DePontieu2014}.  We focus  this study on a prominence observed on July 17, 2014, with{ an apparent} helical structure (Section 2). A few days before, this structure  was a north-south oriented  filament, %with a prominent footpoint in its central part (
 with some extent in the east-west direction (Figure \ref{Meudon}).   The IRIS and SOT fields of view were  centred  on  a section of the filament as it crossed the limb (Section 2). 
 The prominence  has an anti-symmetric Doppler pattern in the \ion{Mg}{II} k line observed with  the IRIS spectrograph (Section 3). The reconstruction of the vector velocity field  in 3D is possible by combining  high spatial and temporal resolution IRIS SJI images in \ion{Mg}{II} k and IRIS spectra. { With the 3D reconstruction we demonstrate that the apparently helical  prominence  consists of a horizontal magnetic structure with no real twist (Section 4)}.
 %________________________________________________________________
  
  %Figure 3 
  
  \begin{figure*}
   \centering
   \includegraphics[width=0.48\textwidth]{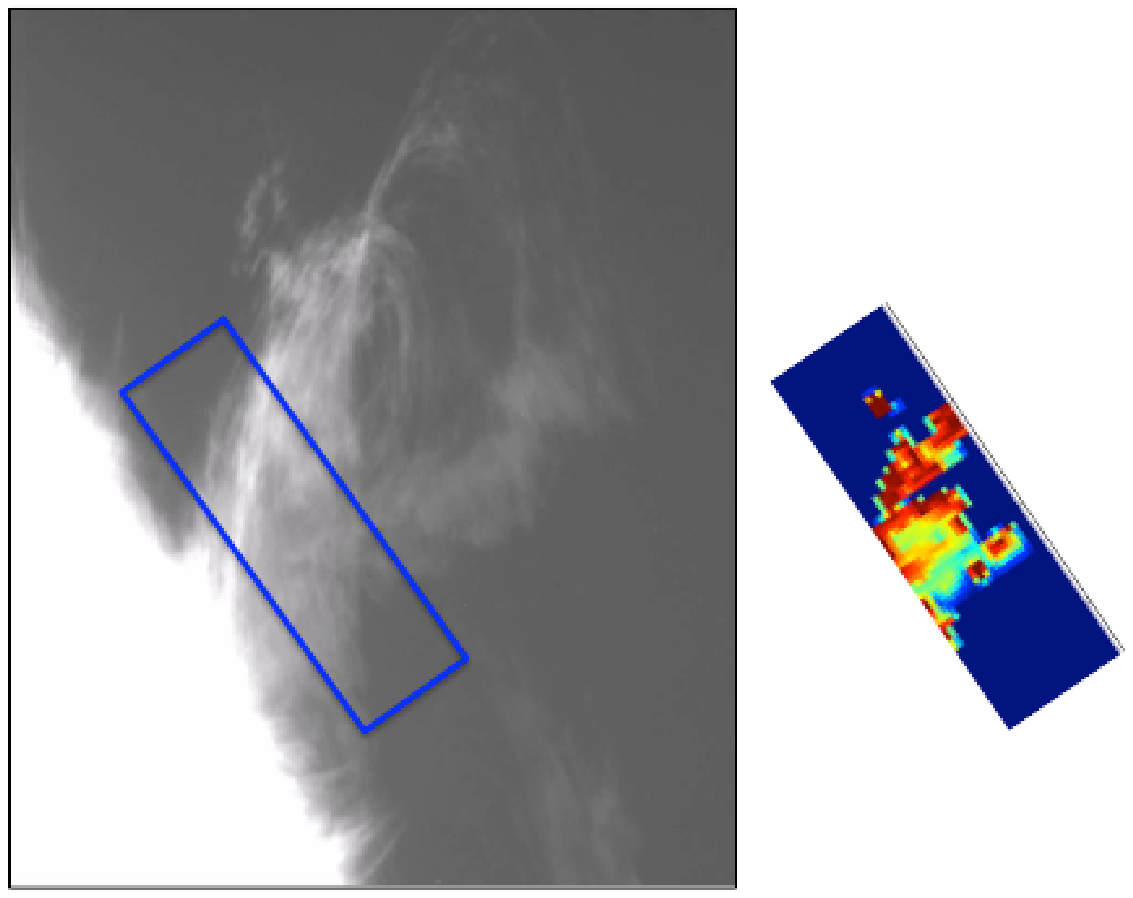}
     \includegraphics[width=0.48\textwidth]{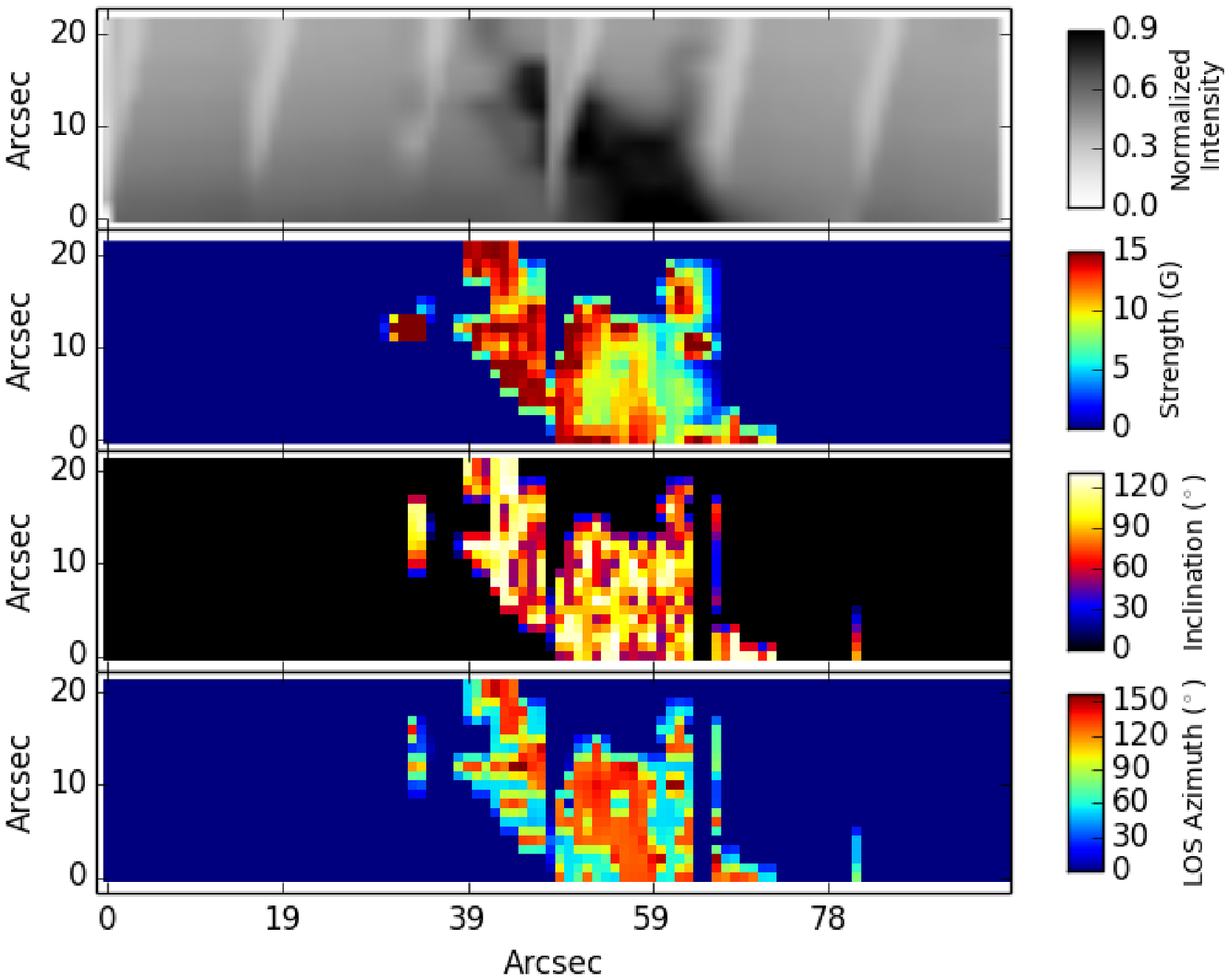}
   \caption{Prominence observed on July 17, 2014  by SOT in Ca II  at 09:02 UT and  its magnetic field strength  from THEMIS   at 09:29 UT overlaid by  a  box, which represents   the field of view of THEMIS observations presented in the right panels: (from top to bottom) intensity in the \ion{He}{I} D$_3$ line, magnetic field strength, magnetic field inclination{ with} respect to the local vertical and magnetic field azimuth{ with} respect to the line of sight. In the rotation of the field of view of THEMIS,  the limb is along the x-axis.   }
\label{SOT_THEMIS}
\end{figure*}
\label{sec:intro}

%Figure 4
     \begin{figure}
   \centering
  \includegraphics[width=0.48\textwidth]{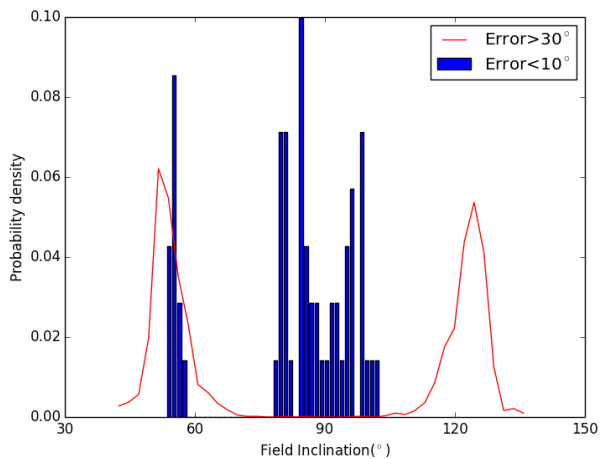}
   \caption{Histogram of the magnetic field inclination{ with} respect to the line of sight measured in the prominence on July 17, 2014.}
   \label{histo} 
   \end{figure}
 
\section{Observation of a helical prominence}
\label{sec:observations}

%\subsection{Instruments}
\label{ssec:inst}

\subsection{IRIS}
\label{sssec:iris}
IRIS performed a 16-step coarse raster observation from 08:40:07 UT  to 12:01:44 UT on July 17, 2014. The pointing of the telescope was (850\arcsec, 454\arcsec), with a  spatial pixel size of 0.167\arcsec. The raster cadence of the spectral observation in
both the near ultraviolet (NUV, 2783 to 2834 \AA) and far ultraviolet (FUV, 1332-1348 \AA\ and 1390-1406 \AA)
wavelength bands was 86 seconds. Exposure time was 5.4 seconds per slit position. 140 scan images, each covering 30\arcsec$\times$119\arcsec,  can be reconstructed   using the 16  spectra obtained during the raster process of the region with a step of 2 \arcsec.
Slit-jaw images (SJI) in the broad band  filters 
(2796 \AA\ and 1330 \AA) were taken  simultaneously with a cadence of 11 seconds. 
The filed of view (FOV) of the SJI was 
119\arcsec$\times$119\arcsec.  Calibrated level 2 data is used for this analysis, with dark current subtraction, flat field correction, and geometrical correction having all been taken into account \citep{DePontieu2014}.

We mainly use the \ion{Mg}{II} k 2796.35~\AA\ { line}
%and \ion{Mg}{II} h 2803.5~\AA\ lines, 
along with the slit-jaw images in the 2796~\AA\ filter for this study. The \ion{Mg}{II} k line is 
%and k lines are 
formed at chromospheric plasma temperatures ($\sim$10$^4$ K). The SJI 2796 \AA\ filter  is dominated by emission from the \ion{Mg}{II} k line.
%while emission in the 1330~\AA\ slit jaw is an integration of the FUV emission from within a range of about 40~\AA, including the total emission of two \ion{Si}{IV} lines and UV continuum. 
%formed in the lower chromosphere and the Si IV 1402 and1393~\AA\ lines formed in the prominence transition region (PCTR). 
The co-alignment between the optical channels is achieved by comparing the positions of horizontal fiducial lines.

 %__________________________________________________________________

 %figure 5 
      
     \begin{figure*}
   \centering
    \includegraphics[width=0.9\textwidth]{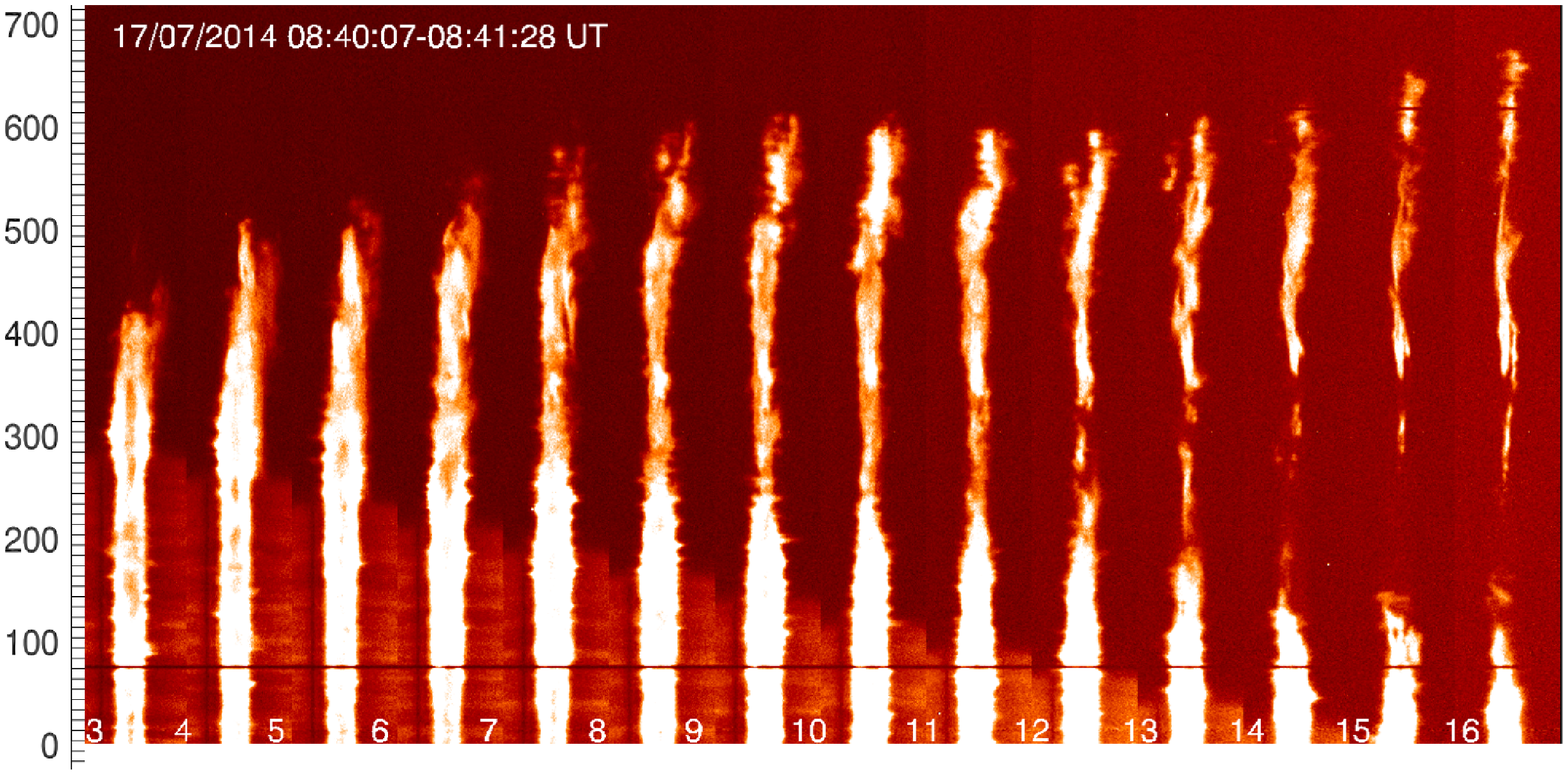}
    \caption{IRIS \ion{Mg}{II} k spectra   during the  first scan between 08:40:07 and 08:41:28 UT. Only the spectra crossing the prominence are shown (3 to 16).
    % In the left top corner is the zoomed image of the knot in the SLJ, in the right bottom corner is the spectra of the Mg II k line.
   % The numbers on the left indicate the number of pixels along the slits (the size of  pixel is O.167 arc sec). The 16 spectra are distant of 2 arc sec in the x direction. The dashed inclined line represents the limb.The green dashed horizontal lines are drawn to follow the  left  leg  and the top of the ellipses. 
     The left column represents the number of the pixels along the slit with a unit equal to the pixel size (0.167\arcsec). 
     %The limb  position is approximatively  represented  by a white dashed line. 
     The number  of each position of the slit is indicated in the spectra at the bottom of the image. 
     %The green dashed lines give some references for the top of the ellipses for   spectra  $>$ 7.
    }
   \label{iris} 
   \end{figure*}

%Figure 6 
 
  \begin{figure*}
   \centering
  \includegraphics[width=0.8\hsize]{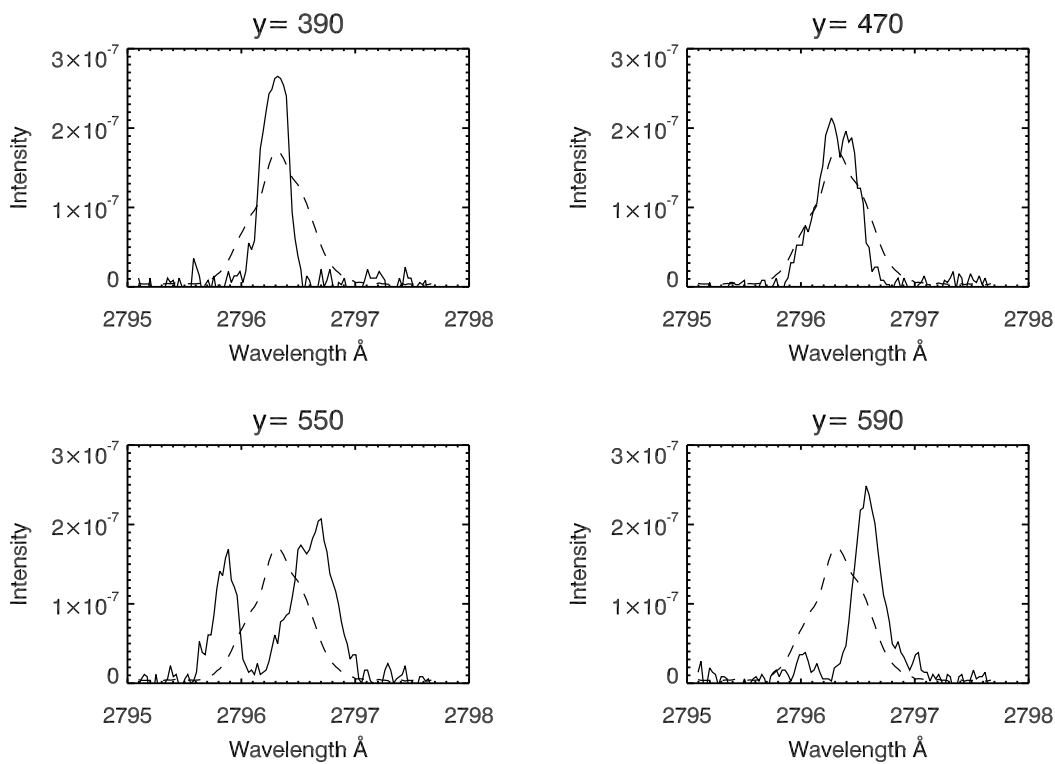}
      \caption{Examples of \ion{Mg}{II} k line profiles along slit 13 at 08:41:12 UT during   the first scan of the tornado.  
      The dashed profile is a mean profile, averaged along the slit. The{ graph titles} represent the position of the  pixel along the slit (see Figure \ref{iris}). The{ intensity} unit is in erg s$^{-1}$ sr$^{-1}$ m$^{-2}$ Hz.$^{-1}$.
    }  
         \label{profile}
   \end{figure*}

%Figure 7
 
     \begin{figure*}
   \centering
  \includegraphics[width=0.8\hsize]{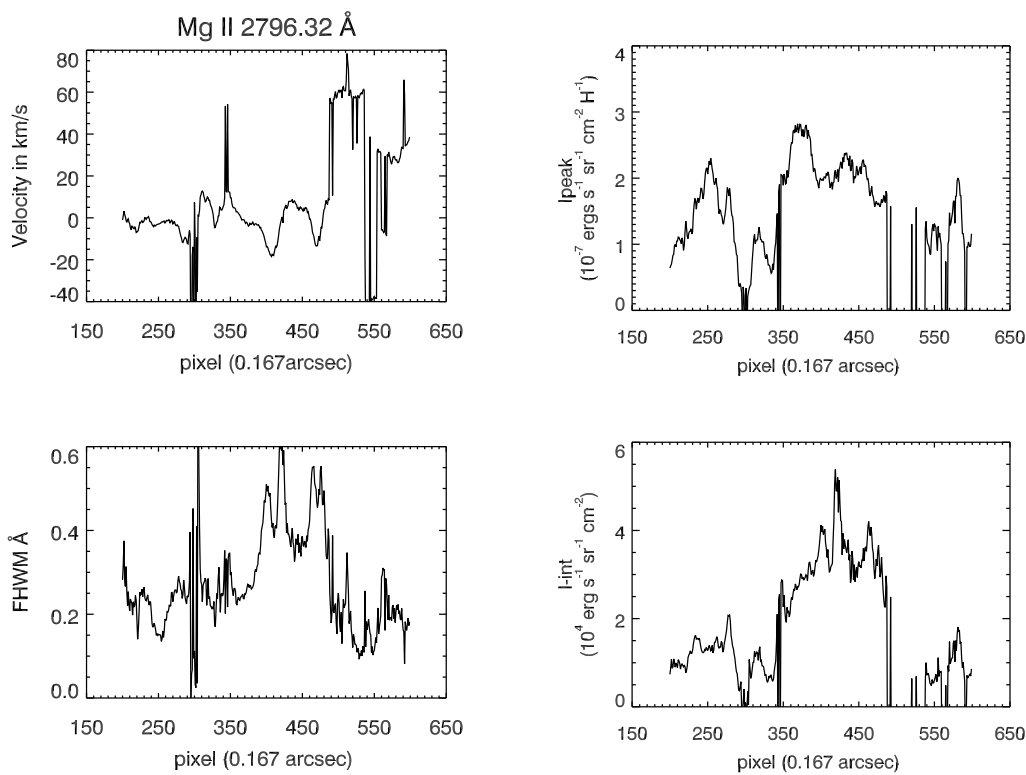}
  \caption{Doppler shift, peak intensity, total intensity, and FHMW of the \ion{Mg}{II}  k profiles along slit  13    at 08:41:12 UT  during the first scan of  the tornado.  
  % The origin of the x axis is pixel y= 250 in  Figure \ref{iris}.
  }
   % in  the code  slit =12 
   \label{FHMW} 
   \end{figure*}

 %Figure 8

\begin{figure}
   \centering
 \includegraphics[width=0.6\hsize]{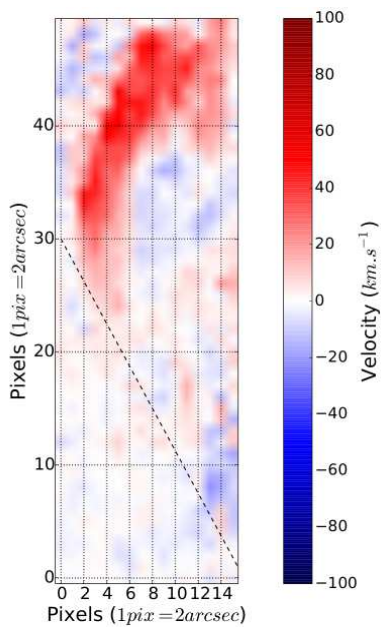}
 \caption{Doppler shift map of  the prominence observed on July 17, 2014  between 08:40:07 and 08:41:34  UT with IRIS. The  Doppler shifts have been computed from the gravity centre of the \ion{Mg}{II} k line profiles. The pixel{ size} in the y direction has been degraded to 2\arcsec\ to have a square pixel.  { The colour bar shows Doppler shift in km s$^{-1}$.} { The dashed line approximately indicates the limb.}}
         \label{Doppler}
   \end{figure}
   
%Figure 9
\begin{figure*}[ht!]
\centering
\includegraphics[width=0.9\textwidth]{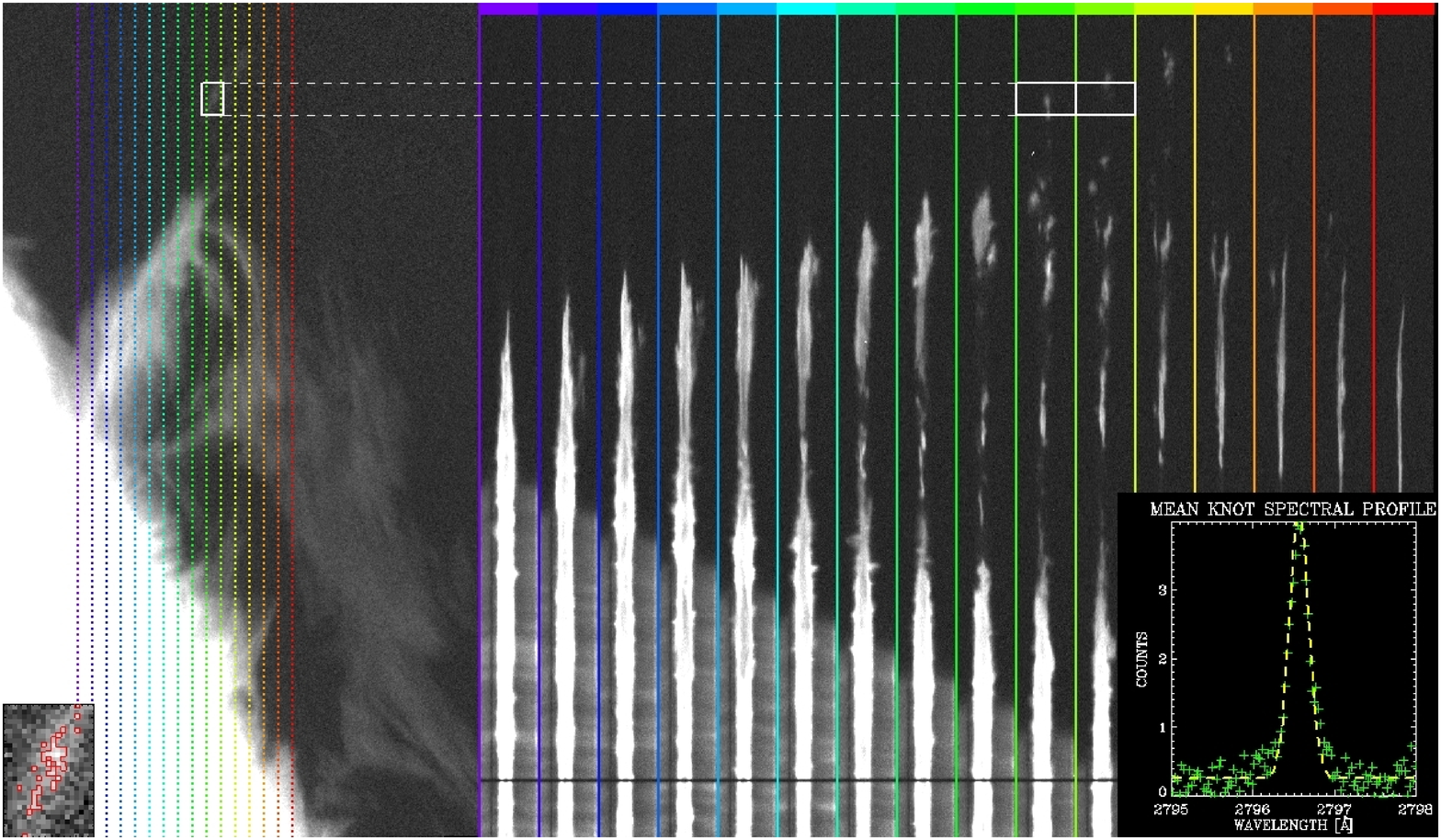}
\caption{Slit-jaw image and corresponding spectra taken during one scan between 10:15:10 and 10:16:29 UT (animated image is online, see Movie 2). Dotted vertical lines{ indicate} slit position in the FOV. Corresponding spectra are colour coded. White rectangle on the slit-jaw image shows knot area, which is magnified in the bottom-left corner. Red lines show 99\% isophote. White rectangle on the spectra (top-right){ indicates area of the} spectrum{ which is} taken for further analysis. }
\label{XXX}
\end{figure*}

\subsection{Hinode/SOT}
\label{sssec:sot}
The Hinode/SOT telescope  consists of a 50 cm diffraction-limited Gregorian telescope and a Focal Plane Package including the narrowband filtergraph (NFI), broadband filtergraph (BFI), the Stokes Spectro-Polarimeter, and Correlation Tracker (CT). For this study, images were taken with a 16 second cadence in the \ion{Ca}{II} H line at 3968.5~\AA\ using the BFI  with a 3 \AA\ bandpass between 08:04:30 UT and  09:02:54 UT. The \ion{Ca}{II} images have a pixel size of 0.109\arcsec\   and an exposure time of 1.22 s, with a field of view of 112\arcsec$\times$112\arcsec.

 Figure \ref{Meudon} (right panel) and the  Movie 1 (animated movie of Figure 1)  show the helical structure of the prominence observed  with SOT. The observing times of IRIS and SOT overlap{ for} only 20 minutes. The  spatial scale of SOT  (0.109\arcsec) is better than{ that of the}  IRIS SJI  (0.167\arcsec\ $\times$ 0.167\arcsec), therefore  very fine structures are resolved.
%During one hour we see  in the SOT movie, knots    moving up and down  along  expanding, and  then shrinking helical structures.   
{ In the SOT movie we see{ some small fragments of the prominence, that we will call ``knots'',} moving up and down along expanding, and  then contracting, helical structures over the course of an hour. }
At the beginning of  Movie 1, the helical structure rises, reaches the top of the field of view, and progressively the  height of the loops is reduced.
 This is visible if we compare the three images of SOT  at three different times within the same field of view  in Figures \ref{Meudon}, \ref{time}, and \ref{SOT_THEMIS}.

 Figure \ref{time} { (right panel) shows the} time distance diagram obtained along a path following  one ellipse of the helical structure from its  right leg to its  left leg  after having co-aligned all the frames of SOT.  The  right leg (between 0\arcsec\ and 20\arcsec) is moving toward the right and  the motions of the knots cannot be{ seen} because they are going out of the path, but the knots along the right  leg (between 40\arcsec\ and 90\arcsec) show transverse velocities from 10 km s$^{-1}$ to 70 km s$^{-1}$.  They are all descending in the same direction indicating  the existence of many parallel  threads.

\subsection{THEMIS spectropolarimetry}
%\label{ssec:themis}

For several campaigns, the French telescope T\'elescope H\'eliographique pour l?Etude du Magn\'etisme et des Instabilit\'es Solaires (THEMIS) in the Canary Islands \citep{Rayrole2000} with the  MulTi-Raies (MTR) mode  has adopted an almost fixed setup for the observation and measurement of magnetic fields in prominences. The large collection of prominences observed has resulted in several publications 
%where this setup has been described in detail 
 \citep{Schmieder2013,Schmieder2014,Levens2016a}.
In the observations{ used in} the present work, the slit of the MTR spectrograph, oriented parallel  to the limb, 
was scanning mainly the base of the prominence (see the blue box in Figure \ref{SOT_THEMIS} left panel). The observations consist of three  successive rasters{ of 20 positions with 2\arcsec\ between slit positions }
starting at 09:29 UT, 10:26 UT, and 11:20 UT. The pixel size along the slit was of 0.2\arcsec, but seeing conditions and the long exposure times required for polarimetry  reduce the resolution to around 1\arcsec.

The acquisition of a full raster{ takes} less than one hour.   {THEMIS  polarization analysis is based  on  a beamsplitter polarimeter. A grid-like mask   is used to split the field of view and leave place, at intervals of 15.5 arcsec, for the secondary beam originating in the polarimeter beamsplitter (more details  available in \citet{Schmieder2013})}.
% used by the polarimeter, 
The images are obtained by two successive displacements  of the grid  along the slit to cover the full spectra of the four Stokes parameters (I, Q, U, and V) in the doublet of the \ion{He}{I} D$_3$ 5876 \AA\ line. This is the reason behind  the{ light grey}  vertical bars in the images.{ The bars are not rectangular due to some shift in the instrument during the observation (Figure \ref{SOT_THEMIS} top left panel).}

As in previously cited studies of prominences using this setup,  raw data  was reduced using the DeepStokes procedure \citep{LAARMSDG09} and the resulting Stokes profiles were fed to an inversion code based on Principal Component Analysis  and able to identify both the Zeeman and Hanle effects in the Stokes profiles \citep{LAC02,Casini03}.{ These} codes rely on a 
database of pre-computed profiles (around 90000 for the database used here) among which the solution is found. The database is dense enough for the solution to fit the observations up to noise levels if the observed profile can be explained by a single vector magnetic field per pixel in the absence of radiative transfer. Error bars for each of the physical parameters retrieved are determined by performing statistics on all other models which are similar to the observed profile, but not as similar as that which is selected as the solution.  Errors can have two meanings: random errors are due to the combined effect of noise in the data and the coarseness of the database. They are acceptably small. For example, random errors in the inclination of the magnetic field are known to be on  the order of 10$^\circ$. Larger errors are the result of either ambiguities (for example the 180$^\circ$ ambiguity in the azimuth of the field) or, more often, are{ caused by} observed profiles that cannot be explained by the simple model of a single vector field per pixel.

%______________________________________________________________

%\subsection{Morphology of the prominence}
%\label{ssec:morphology}

%\section{Physical parameters}
%\label{sec:parameters}

%\subsection{Magnetic field vector}

Figure \ref{SOT_THEMIS} (right panels)  presents the maps obtained after the inversion of the Stokes parameters recorded in the \ion{He}{I} D$_3$ line { (from top to bottom)}: Intensity, magnetic field strength, inclination { with} respect to the local vertical, and azimuth{ with} respect to the line of sight. The origin angle of inclination is the local vertical and the origin of the azimuth is the line of sight (LOS) in a plane containing the LOS and the local vertical. We see that the brightest parts of the prominence have a mean inclination of 90$^\circ$, which means that the magnetic field in these parts is horizontal. However, there is a large dispersion of the values ($\pm$ 30$^\circ$) from one pixel to the next in  the prominence. The 60$^\circ$ inclination pixels are mainly in the centre of the prominence between the two different legs.
%%%%Figure of the first raster should be added.

 The histogram of the inclination for all the points in the prominence  
 shows that{  the quality of the inversion is not as good as for former prominences \citep{Levens2016b}. There are only a few  points in the  peak,} centred on 90$^\circ$, indicating an horizontal direction (Figure \ref{histo}). Those results have small error bars of less than 10$^\circ$ which, as said above, correspond to random errors mostly due to noise in the data. We are confident that these horizontal magnetic fields in the prominence are correctly measured.  { Most of the points belong to  two secondary peaks   at} around 50$^\circ$ and 120$^\circ$, but they show    large error bars ($>$ 30$^\circ$). A few points around 
 55$^\circ$ have an error bar less than 10$^\circ$.  { 
 A first conclusion would be that the inclination is 60 or 110 degrees and  the  Principal Component Analysis  (PCA) inversor code cannot decide  what the best solution is  because of the 90 $^\circ$   ambiguity. 
 However, in   another 
  prominence observed by IRIS  where the histograms showed similar distributions of  values, \citet{Schmieder2014} related this distribution  to  the  fact that the prominence was very dynamic. A numerical test confirmed that  when the polarized  profiles were due to  the addition of an horizontal field plus a turbulent field, the resulting profile  would yield  60-110 solution with 30 $^\circ$  errors.  Since this paper, many prominences  \citep{Lopez2015} have shown similar histograms and we propose the same   interpretation  for the present case with}
 the superposition of two magnetic fields inside the pixel: an  horizontal  background field set inside a turbulent ambient field.
  We{ note} that these pixels  with large errors are mainly between the two legs of the{ elliptical loops} where no structures could really be resolved.
%The  inclinations of 30 degrees and 150 degrees, on the other hand have  also been  observed in other tornadoes (Schmieder et al 2015 IAU305). We concluded from this that the magnetic field model of one field per pixel used by the PCA inversion code 
%i s not valid. We can definitely exclude a vertical field as solution, which would have been correctly inverted. Several possibilities mixing several magnetic
% components as in the case of the prominence have to be  explored.

%______________________________________________________________
%______________________________________________________________

\section{Spectral analysis of IRIS}
\label{ssec:iris_analysis}

\subsection{Doppler shifts}
\label{sssec:doppler_shifts}

Figure \ref{iris} shows an example of a  set of  \ion{Mg}{II} k spectra obtained during the first raster. The left column indicates the pixel number along the slit.  We note the presence of reversed profiles on the disk.
% below the dashed white line corresponding to the chromosphere. % PJL: I SEE NO DASHED WHITE LINE IN FIGURE 5...?
Spectra 3 to 6 show a dense part of the prominence,{ while} spectra 7 to 16{ show}  a less dense part where  elliptical loops can be distinguished. The profiles of the \ion{Mg}{II} k line{ spectra from the} helical{ part of the} prominence are mainly non-reversed.
%(see the upper part of the spectra 10 to 16  limited by the green lines of Figure \ref{iris}).
The spectra  in the upper part 
%above the green lines 
are very twisted indicating high Doppler shifts.
% and a SJI overlaid by the 16 slits  of one raster (movie 1 on line).
%The spectra have been calibrated. 
Examples of  profiles are shown in Figure \ref{profile}.
The reference profile in these plots is a mean profile obtained by averaging  the profiles of pixels above the limb along one slit. 
{ We note that there are a number of different profiles in this prominence:} some profiles are narrow (y=390),  some profiles are reversed (y=470), some profiles  have double components (y=550), indicating the presence of two or more structures along the line of sight,{ and in some locations the entire profile is Doppler} shifted (y=590).

Different methods have been tested to compute the Doppler shifts.
In a first attempt, the profiles { are} fitted{ using} Gaussian functions{ which are then used} to derive the main characteristics of the prominence plasma. The Full  Width at Half Maximum (FWHM)  is around 0.2 \AA\ and  the profiles are commonly narrow in prominences, as was{ the case} in \citet{Schmieder2014}. However, there are exceptions where the FHMW reaches 0.4 or 0.6 \AA.  In these cases it appears that the profiles are certainly a combination of 
different structures along the LOS with different velocities. Assuming Gaussian profiles for the IRIS \ion{Mg}{II} lines, we return relatively small line-of-sight velocities ($\pm$ 5 kms$^{-1}$) for wide profiles.
For  narrow profiles, higher Doppler shifts are measured, reaching 60 km s$^{-1}$. These points correspond to the  highest points along the slit and more or less   the top of the prominence (e.g. profile at  y=590 in Figure \ref{profile}).{ Such points belong to the knots that have been followed along the elliptical structures (see Section 4)}.

%The Doppler shifts using the assumption of gaussian line profiles lead to small velocities (+/-5 km/s).  The pattern obtained for the velocities follow the intensity pattern of the loops between the two columns with alternatively red and blue shifts (Figure maps of dopplershifts).  The evolution versus time is slow . In one hour we see the blueshift on one side of the loop has reached the other  end of the loop.   Working on the moments of the line profiles, the pattern is unchanged. The values are of the same order. 

\subsection{Doppler pattern}

We analyse the \ion{Mg}{II} spectra of  the first  raster, beginning at 08:40:07 UT  and ending at 08:41:24 UT,  along  the 16 slit positions. 
As some  profiles are reversed  (either due to absorption or to the presence of  different structures along the line of sight) it{ is more useful} to 
 compute Doppler shifts by using the gravity centre of   each profile  than to  fit the profile  with a Gaussian curve.
 The spatial resolution is degraded in the y direction  to 2\arcsec, which corresponds to the{ step size in x}.
%a technics separating the blue part of the profile from the red part of the profile 
We  obtain the Doppler pattern of the prominence 
%considering independently  the blue part and the red part,  then overlapping both images 
(Figure \ref{Doppler}).  The resulting  Doppler map   shows  an anti symmetric pattern with strong redshift in the left part of the prominence and blue shift in the right part, similar to  those patterns found with the spectroscopic data of EIS  in \ion{Fe}{XII} at 195 \AA. These patterns  could suggest the existence of rotation around the axis of the prominence,{ an observational  known as a} ``tornado''.

\section{3D  trajectory reconstruction}

The  SOT movie (Movie 1 and Figure \ref{Meudon} ) and the IRIS  SJI movie (Movie 2 and Figure \ref{XXX}) show knots following helical trajectories in the plane of the sky.
Even{ though} the IRIS movie has a slightly lower spatial resolution, it has been possible to follow manually  a few of{ the} knots using the IRIS 2976 \AA\ SJI{ along with}   the simultaneous  \ion{Mg}{II} k spectra.  In the set of  slit-jaw images, we detected several prominence knots. Knots are fragments of prominence plasma, which are detached{ from the main prominence} and move independently along magnetic field lines. They{ are} approximately $10^3-10^4$ km in diameter and{ usually} circular or elliptic{ in} shape. 

Using the 3D  trajectory reconstruction  method   described in \citet{Zapior2012},  the positions of  15 knots in x and y{ are} computed by polynomial approximations. Only nine knots are shown in  the bottom right panel of Figure \ref{BBB} for clarity. Having analytical expressions for x(t) and y(t), projections of velocity vectors v${_x}$(t) and v${_y}$(t)  are calculated as derivatives of x(t) and y(t).  The Doppler shifts of the IRIS spectra give the third component  along the{ z axis,} which is  oriented away { from} the plane of the sky,  x and  y being  in the plane of the sky (plane of the images).  The reference system{ used here} is not the conventional one.  The y-axis makes  an angle of 60$^\circ$ with the radial direction  and an angle of 10$^\circ$ with the main direction of the{ prominence body}, which is inclined towards the solar surface.
Having the set of slit-jaw images together with the spectra, at least in the part of the prominence image, it{ is} possible to reconstruct true 3D trajectories of knots. 

From the set of observations we{ make} a ``sandwich'' median image from slit-jaw images of a particular scan. In other words, we stack data cubes, which consist of eight slit-jaw images per cube, and we constructed { the} median image, which has a median value for each pixel. This{ gives a} better signal-to-noise (S/N) ratio and even faint knots become visible. One scan lasts about 86 seconds.{ Over this} period,{ the} shift of{ an individual} knot is not substantial.{ However, spectra were taken 2 times as  frequently  (i.e 16 spectral observations during one scan). We present slit positions{ in an} exemplary slit-jaw image in Figure \ref{XXX}.

In the set of consecutive slit-jaw images, we select knots visible in at least several consecutive images. We manually{ select an} area with a knot in{ the} slit-jaw images (white rectangle in Figure \ref{XXX}) and we{ identify  knots} as regions with signal above 99\% of the brightest pixel inside the area, marked with a red isophote (see left bottom corner of Figure \ref{XXX}). From the centre of gravity of the signal inside the isophote, we calculate the position of the analysed knot ($x_n,y_n$) in the plane-of-sky (POS), where $n$ is the scan number, counted from the beginning of the observation of a particular knot.

%Then we found corresponding knot spectrum in the following way.

Then, we{ take the} spectrum corresponding to the nearest slit position (see colour coding in Figure \ref{XXX}). If{ the} selected knot area crosses more than one slit position, we calculate{ the} spectrum as a weighted mean from{ the} corresponding slit positions. We project{ the} region limited by the isophote onto{ the} vertical axis and make a sum of{ the} signal in each row. We construct a vector of accumulated signal from each row in this way. We take into account only pixels from inside the isophote. We normalise the vector, which is then treated as a weight, $W_j$, where $j$ is a row number. Next we calculate the mean weighted spectrum, $S({\lambda})$, given by: $$S({\lambda})=\frac{\sum_j S(\lambda,j) W_j}{\sum_j W_j.}$$ This is then treated as a knot spectrum{ at a} particular time $t_n$.

%figure 10-14
\begin{figure*}[ht!]
\centering
\includegraphics[width=0.9\textwidth]{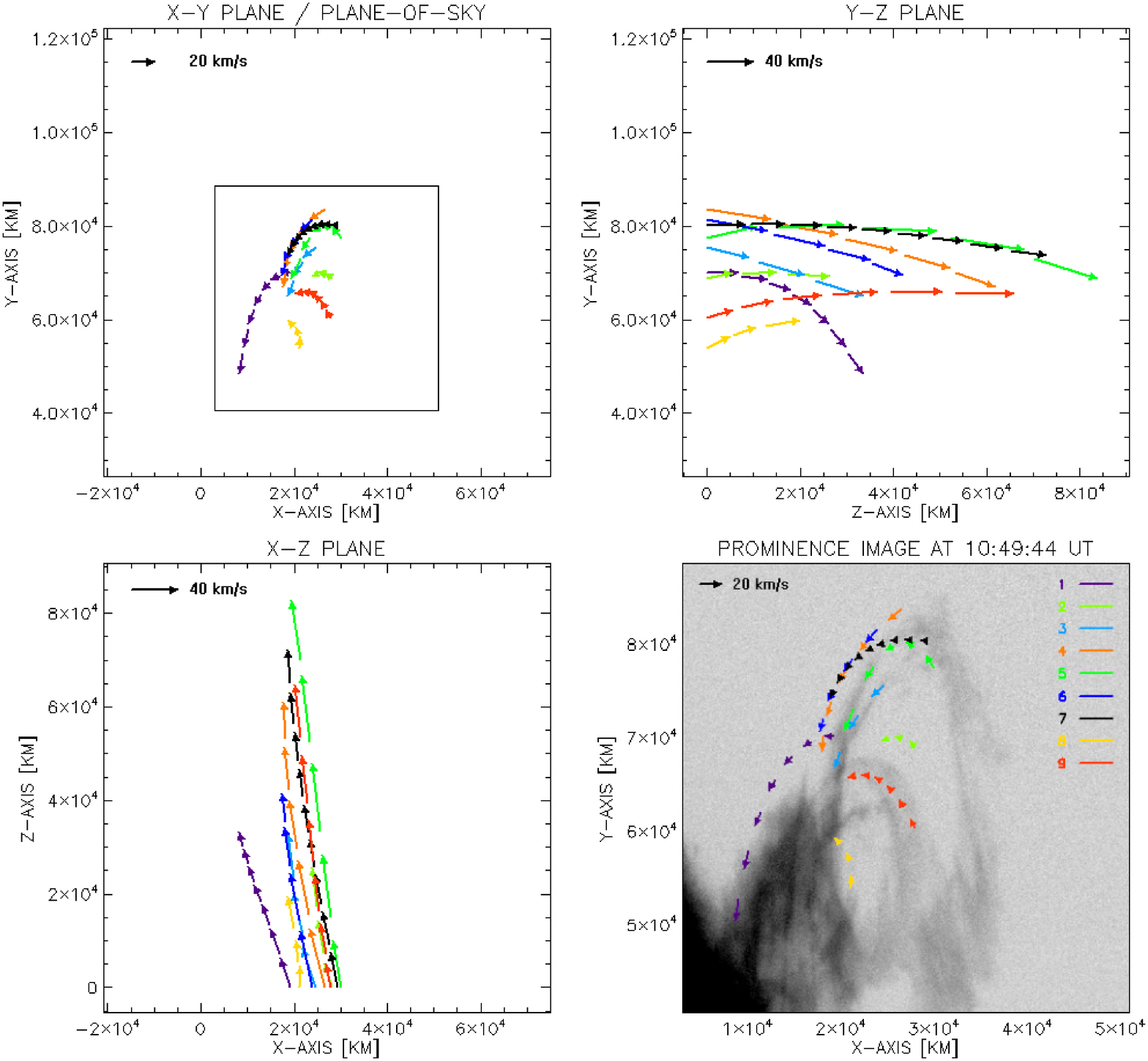}
\caption{Projections of knot trajectories onto perpendicular planes. Arrows give trajectories of particular knots together with projected velocity vector. Axes are oriented as follows: x axis oriented right in the POS along bottom edge of the slit-jaw image, y axis oriented up along left edge of the slit-jaw image, z axis oriented along LOS{ away from} the observer.{ Bottom right panel shows an overplot of the vectors onto a prominence image from 10:49:44 UT} (Movie 4 shows an animation of this). Thread numbers  correspond to numbers in  Table~\ref{TTT}. In the top left corner of each panel, reference velocity vector in the plot scale is drawn. Small rectangle in top left panel shows the area of bottom right panel, which has a different  scale than   the other panels.}
\label{BBB}
\end{figure*}
%Figure 11-10
\begin{figure*}[ht!]
\centering
%\begin{tabular}{cc}
%\includegraphics[width=0.45\textwidth]{nor_profile.jpg}
\includegraphics[width=0.43\textwidth]{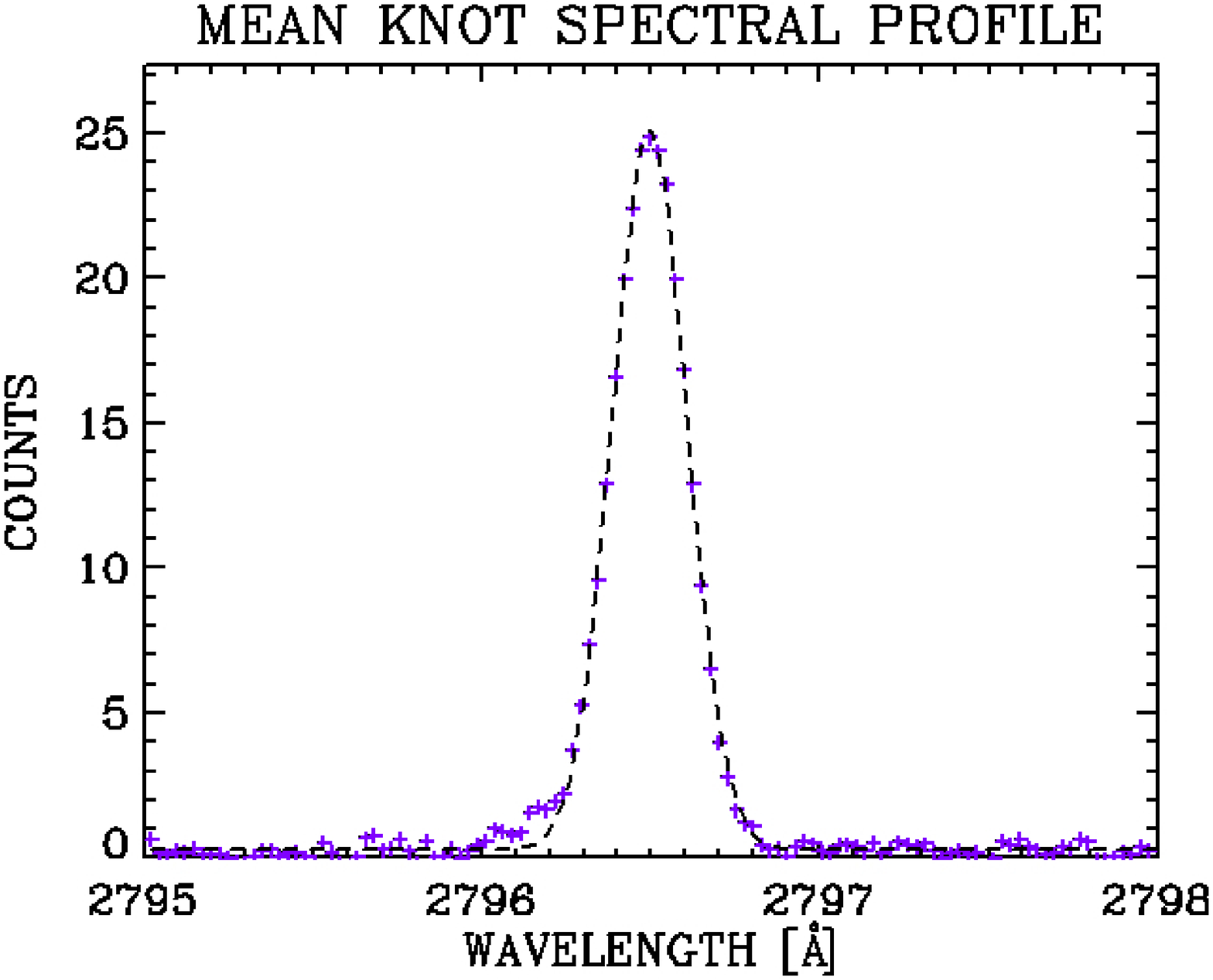}
%&
%\includegraphics[width=0.45\textwidth]{rev_profile.jpg}
\includegraphics[width=0.43\textwidth]{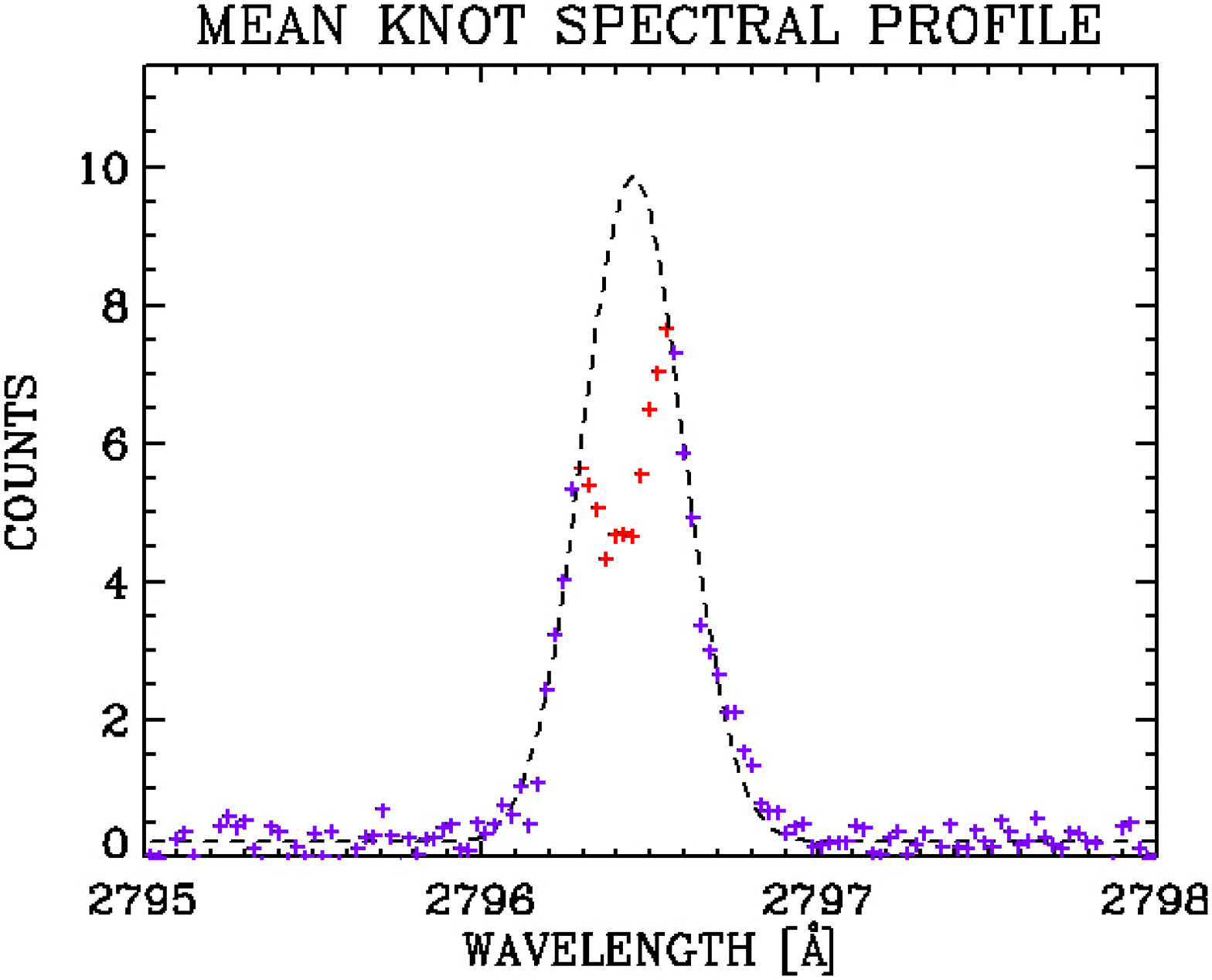}
%\end{tabular}
\caption{(Left panel:) Observed spectral profile (green points) of an analysed knot with regular Gaussian shape. (Right panel:) Spectral profile of an analysed knot with central reversal (red points). Gaussian fitting is performed on green points only. Black dashed lines show fitted Gaussian profile.} 
\label{YYY}
\end{figure*}

For { each} knot spectrum at a given time ($t_n$) we fit a single Gaussian curve 
using a robust, non-linear least squares curve fitting code by \citet{Markwardt2009},{ which is} based on the algorithm by \citet{More1978} (see Figure \ref{YYY}).
The reason for using a single Gaussian is because knots are separate structures visible{ against} the background dark sky. They usually have a small optical thickness. This leads to a single Gaussian profile because there is no other structure along line-of-sight (LOS). For a limited number of spectra ($\sim 25\%$),{ we find centrally reversal profiles}{ due to} the larger optical thickness. In such cases we { remove the} central reversal using an automated procedure. The procedure searches for points in the spectrum with lower signal between{ the} two peaks (see Figure \ref{YYY}, right panel) and{ removes them from the Gaussian fit}. Gaussian { fitting is therefore performed on spectral wings only.}
From { the shift of these Gaussians} we calculate the Doppler velocity. 
For each scan we calculate{ the} mean profile of the quiet photosphere and corresponding residual velocity (see Figure \ref{ZZZ}). Residual velocity, which is caused by{ the} orbital period of the IRIS satellite,{ is} approximated by a cosine curve and is treated as the zero point for correction of Doppler velocity.

%Figure 12-11
\begin{figure}[ht!]
\centering
\includegraphics[width=0.5\textwidth]{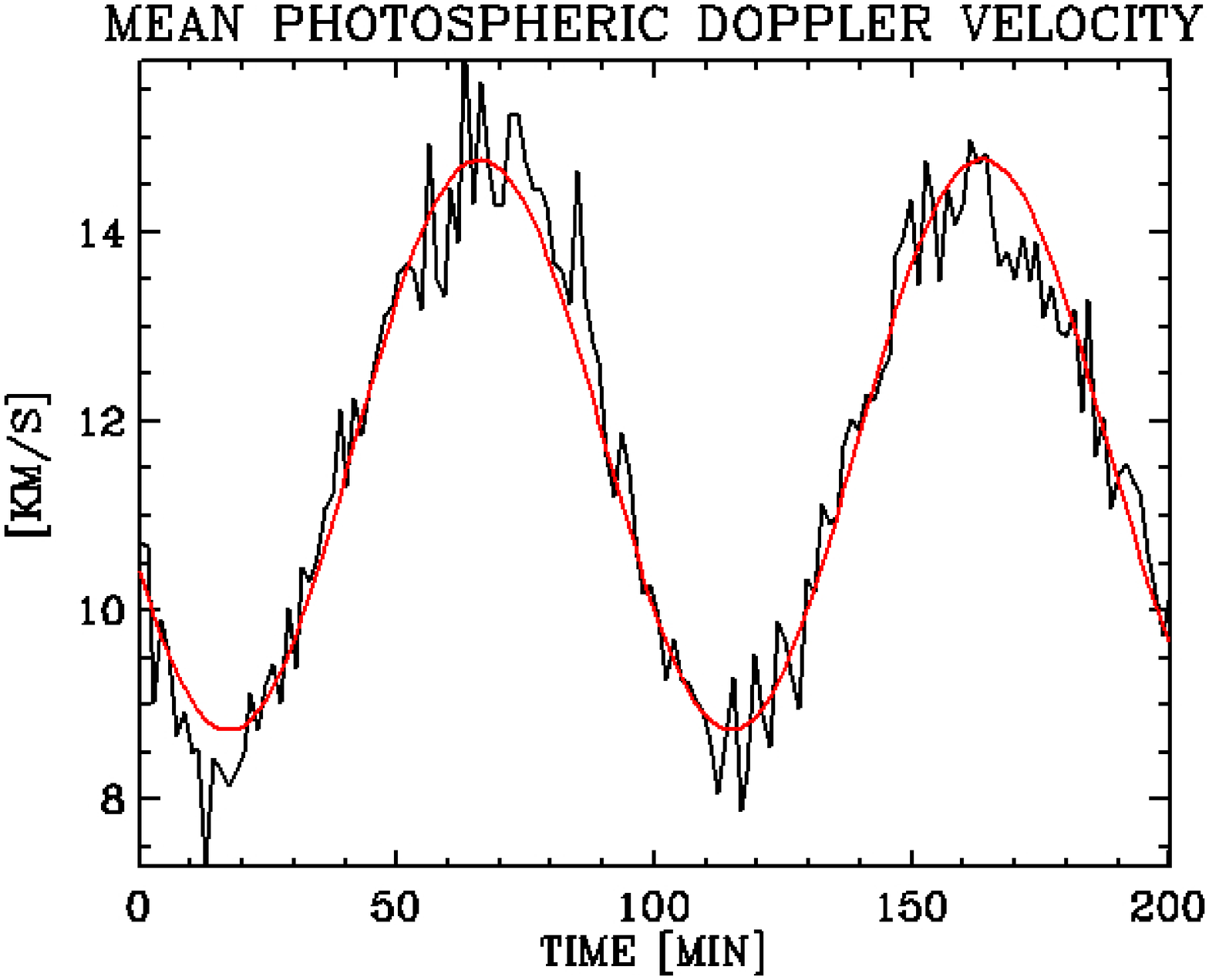}
\caption{Mean photospheric Doppler velocity calculated from{ the} quiet photosphere,{ which is} treated as{ the} zero velocity level.} 
\label{ZZZ}
\end{figure}

For each knot we have a set of data: $t_n$ --  time of observation taken as a time of the first spectrum in a scan, $x(t_n),y(t_n)$ -- POS position of a knot, $v_D(t_n)$ -- Doppler velocity, where $n=1,...,N$ and $N$ is a total number of observations of a particular knot. We then perform{ an} approximation of $x(t_n),y(t_n),v_D(t_n)$ using polynomials. To find a polynomial that properly describes{ the} observed variations, we make use of orthogonal Chebyshev polynomials of the first kind, $T_m(t)$. We successively perform an approximation of the data points $x(t_n),y(t_n),v_D(t_n)$ { using LSE method with a sum of Chebyshev polynomials with coefficients $a_m$ as free parameters} with consecutively higher orders: 
$$ P_m(t)=\sum_{i=0}^m a_m T_m(t), $$
where $P_m$ = fitting polynomial { of the data points}, $T_m$ = Chebyshev polynomial of the $m$ degree. %Approximation was performed.
% using LSE method using robust non-linear least squares curve fitting code by 

% Mor�, J. 1978, "The Levenberg-Marquardt Algorithm: Implementation and Theory," in Numerical Analysis, vol. 630, ed. G. A. Watson (Springer-Verlag: Berlin), p. 105 (DOI: 10.1007/BFb0067690; Link to Springer title listing) 

We calculate { the} \textit{t}-Student statistic $S_m=\frac{\sigma_{a_m}}{a_m}$ for $m=0,1,\dots,10$. Calculation of this  statistic for higher $m$ { values} was unnecessary because knot trajectories are well described by polynomials of low degrees \citep{Larmore1953,Rothschild1955}. { In our case  the maximum polynomial degree used for data point approximation was equal to two (see Table \ref{TTT}).} Calculated values { of} $S_m$ are compared with{ the} critical value of the statistic $t(\alpha,M)$, where $M=N-m-1$, $M$ -- degrees of freedom, $N$ -- number of data points, and $\alpha=0.05$ -- selected significance level. We select{ the} fitting polynomial degree as the highest $m$ for which $S_m > t(\alpha, M)$ occurs. Having established a degree, we{ can then} perform fitting again with{ the} degree equal to $m$, obtaining a  polynomial with { the} correct degree.
{ Rapid changes of the measured LOS velocity were caused by the noise, not by physical changes of the velocity related to the trajectory (see Figure \ref{AAA}). This was caused by the fact that slit positions during scanning were separated. We could not have continuous measurements of the LOS along the trajectory, but only in the slit positions. We decided to approximate LOS data with low degree polynomials in order to avoid over-fitting of the data points. }

As a result of { the} approximation, we have a continuous variation of $x(t), y(t),$ and $v_D(t)$. We then integrate $v_D(t)$ and obtain $\Delta z(t)$ -- the relative shift{ along the} LOS. Having $x(t),y(t),$ and $\Delta z(t)$ (see Figure \ref{AAA}), we have a complete{ spatial} 3D trajectory{ for each} knot. We need to stress here that unless the initial position of the calculated trajectory is not known, values $x(t),y(t),\Delta z(t)$ represent its true 3D geometry. We can arbitrarily shift each knot's trajectory{ along the} $z$ axis, but the 3D shape is conserved (Figure \ref{DDD} and Movie 3). We assume that $z(t_1)=0$. Having analytical variations of $x(t)$, $y(t)$, and $v_D(t)$, calculation of projections of velocity onto perpendicular axes ($v_x,v_y,v_z$), perpendicular planes ($v_{xy},v_{yz},v_{xz}$), spatial velocity ($v_{sp}(t)=\sqrt{v_x(t)^2+v_y(t)^2+v_D(t)^2}$), and local curvature radius ($r_c=v_{sp}(t)^2/a_n(t)$, where $a_n(t)$ - acceleration normal to instantaneous velocity vector) is possible (Figures \ref{BBB}, \ref{VVV}, \ref{CUR}, and Table \ref{TTT}).

All knots have similar kinematic properties. They follow  parallel directions with similar curvature radii (Figure \ref{CUR}). The curvature is calculated locally according to{ the} position of Frenet-Serret formulas, that is, the curvature radius is local in the direction perpendicular to local velocity in the plane of local surface in which{ the} motion occurs. Knots seem to have motions in the plane with no pitch (twist). No helical motions are observed. 3D trajectories of knots are nearly plane curves. Discrepancies of the velocity behaviour  are present because knots are observed in different parts of the prominence body (Figure \ref{VVV}).  Also we do not know the relative position of knot trajectories along LOS, so knots observed close to each other in{ the} POS may{ be more separated than they seem}.

However, we notice some similarity in the behaviour of the { variation of the} velocity norm of a pair of knots: knots  2 and 9 have an increasing velocity, knots 4 and 6 a decreasing velocity. Knots 1 and 7, which are observed{ for} more than 30 minutes, first have a decreasing velocity norm then an increasing velocity norm.  Knots 2 and 9 are on  parallel trajectories  going{ upwards}, knots 4 and 6 are on parallel trajectories going{ downwards}. Their acceleration  behaviour is contrary to 
the { gravitational} force. We conclude that the knots are moving due to gas pressure and{ } not due to gravity.

%Figure 13- 12

\begin{figure*}[ht!]
\centering
\includegraphics[width=0.7\textwidth]{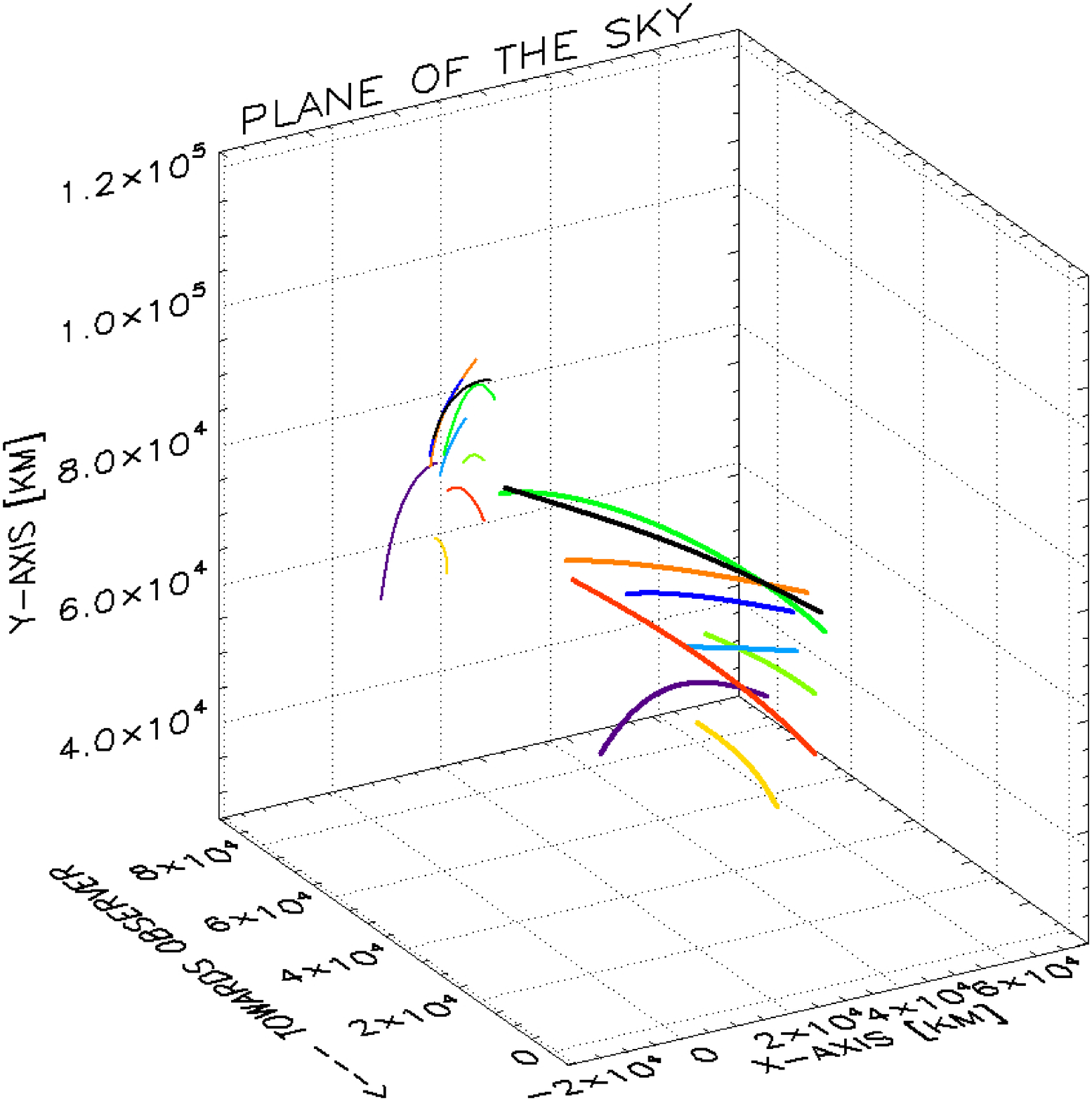}
\caption{Reconstructed 3D knot trajectories in the x, y, and z system (animated cube in Movie 3). X and y are in the plane of the sky, z is oriented{ along the LOS away from the observer}. Thin lines are  projections onto plane of the sky (x,y). Colours correspond to  the colours{ used in} the plots in Figure \ref{BBB}.
 \label{DDD}}
\end{figure*}

%Figure 14-13 
\begin{figure*}[t]
\centering
\includegraphics[width=0.99\textwidth,bb=5 145 540 432, clip]{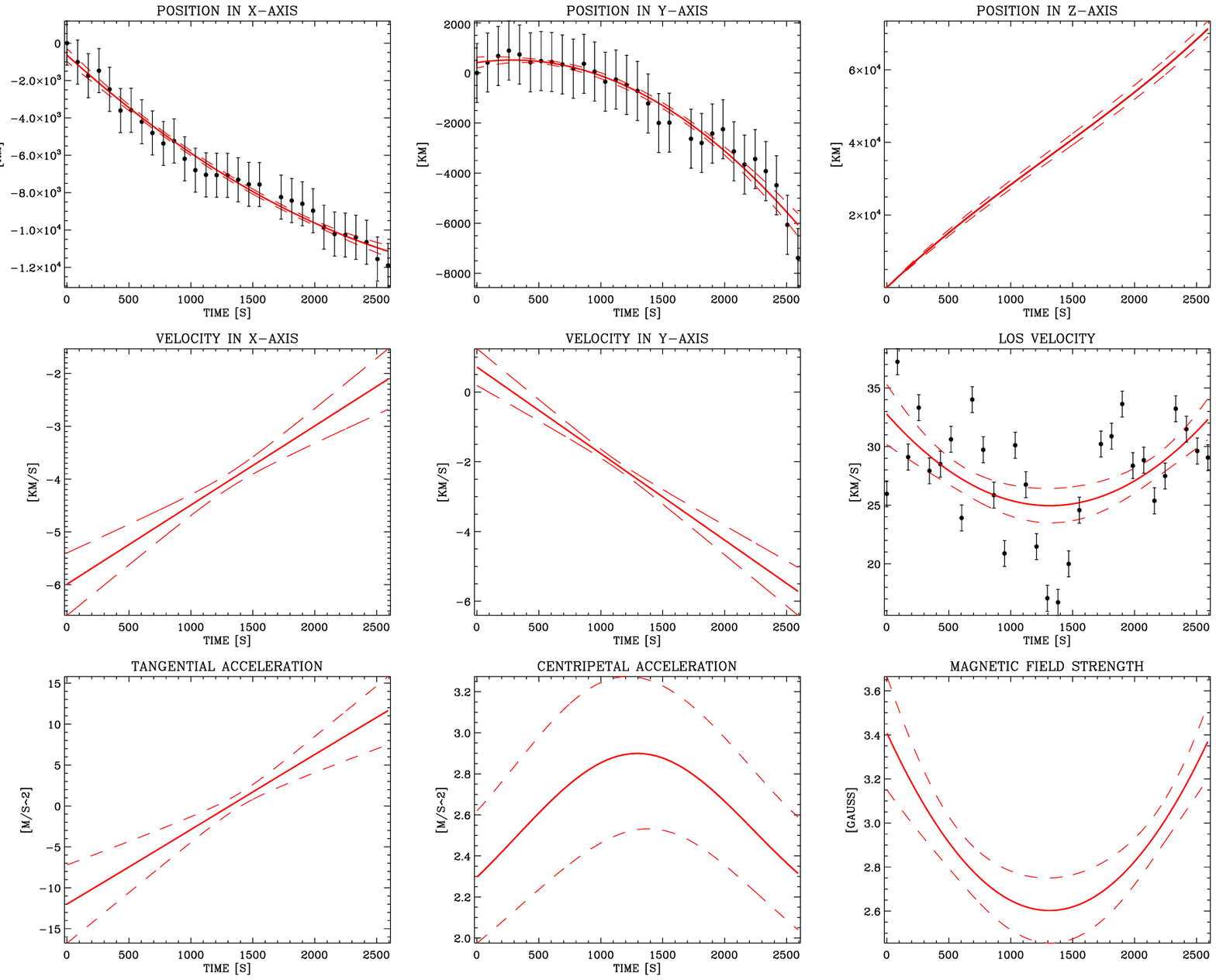}
\caption{Kinematic properties of knot number 7. Black points with error bars represent data points $x_n,y_n,v_n$. Red solid lines represent polynomial fits, their derivatives (for $v_x$ and $v_y$), and integral (for $\Delta z$). Dashed lines represent fitting errors estimated using bootstrap resampling (see e.g. \citet{Andrae2010}). }
\label{AAA}
\end{figure*}

\begin{table*}[t]
\caption{Basic information { about} observed knots. $N$ -- number of observations of the knot, $t_1, t_N$ -- the first and the last time of observation of the knot, $v_{xy}$ --  minimum and maximum values of projection of knot velocity onto plane of the sky, $v_{sp}$ -- minimum and maximum values of the spatial knot velocity, $r_c$ -- minimum and maximum values of the curvature radius along the trajectory, { $m_x$, $m_y$, $m_{v_D}$ -- polynomial degree used for approximating the data points (position on $x$ and $y$ axes and LOS velocity)}.}
\centering
\begin{tabular}{c|c|c|c|c|c|c|c|c|c}
Knot & $N$ & $t_1$ & $t_N$ & $v_{xy}$ & $v_{sp}$ & $r_c$ & \multicolumn{3}{c}{Polynomial degree}\\
number &&&&[km s$^{-1}$] &[km s$^{-1}$] & $10^3$ [km] & $m_x$ & $m_y$ & $m_{v_D}$\\
\hline
1 & 22 & 08:43:00 & 09:13:15 &  8-19  &        19-29 & 30-96    &2&2&2\\
2 & 10 & 10:05:06 & 10:18:03 &  5-8    &      23-41  & 28-162   &2&2&1\\
3 & 8  & 10:32:27 & 10:42:32 &  14-15   &       39-44 & 111-152         &2&2&1\\
4 & 14 & 11:11:20 & 11:32:30 &  13-16    &      36-57 & 110-440 &2&2&1\\
5 & 16 & 09:43:29 & 10:05:06 &  7-22      &    46-65 & 53-168   &1&2&2\\
6 & 12 & 11:58:32 & 11:34:23 &  11-15      &    27-54 & 68-435          &2&2&1\\
7 & 30 & 11:17:06 & 12:00:18 &  5-6         & 25-33 & 61-132    &2&2&2\\
8 & 11 & 11:40:08 & 11:54:32 &  7-11         & 21-40 & 27-104   &2&2&2\\
9 & 19 & 09:19:08 & 09:04:56 &  5-9           &24-51 & 43-422   &2&2&1\\
\hline
\end{tabular}

\label{TTT}
\end{table*}

%figure 15
\begin{figure*}[ht!]
\centering
\includegraphics[width=0.9\textwidth, bb=0 300 338 579,clip]{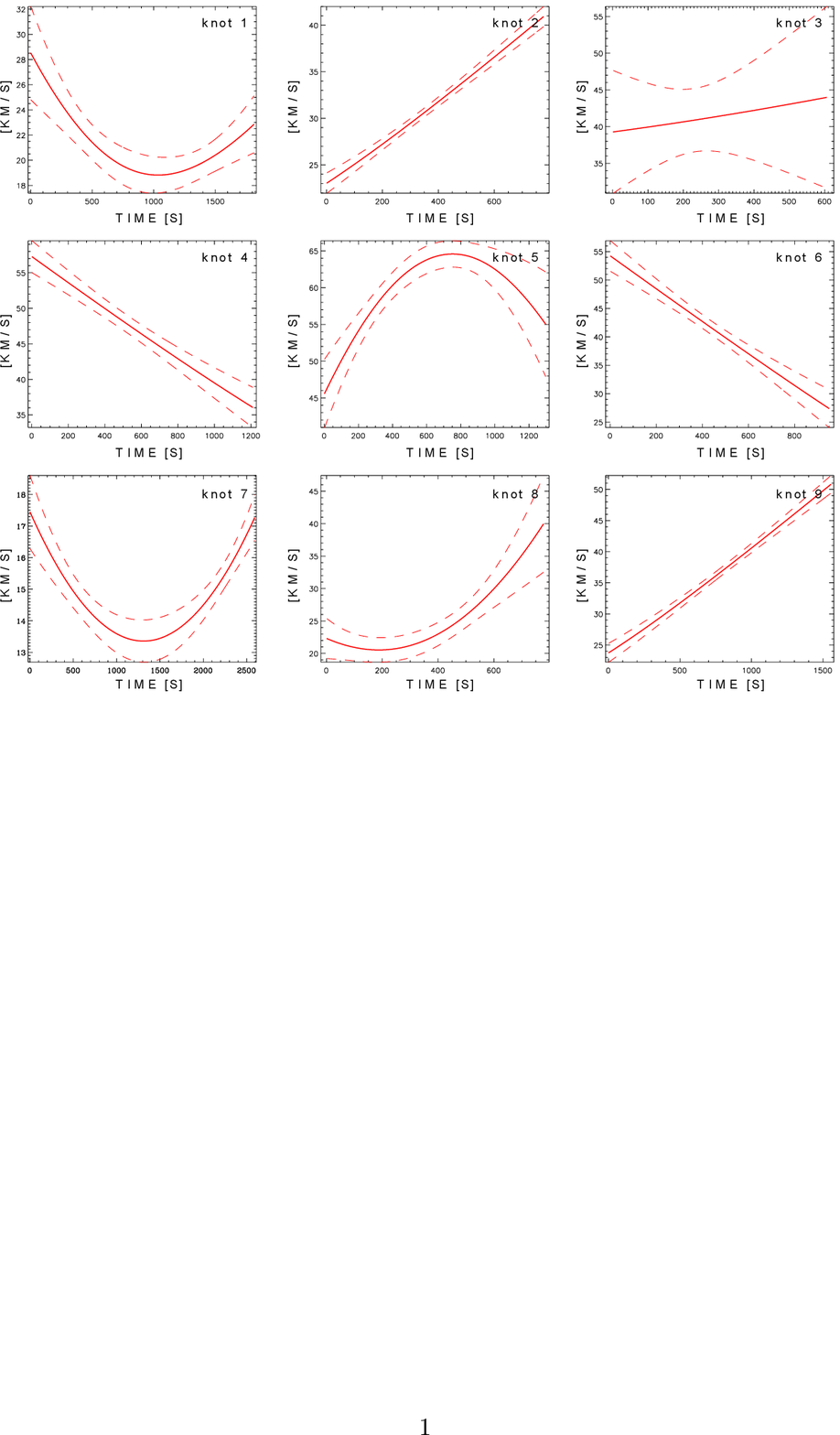}
\caption{Velocity  norm (in km s$^{-1}$) of all observed knots in the space versus time. The horizontal axes are scaled in seconds from the first time of observation of{ each} knot ($t_1$ - see Table \ref{TTT}). Dashed lines give error estimations from bootstrap resampling method.}
\label{VVV}
\end{figure*}

%figure 16
\begin{figure*}[ht!]
\centering
\includegraphics[width=0.9\textwidth, bb=0 300 340 579,clip]{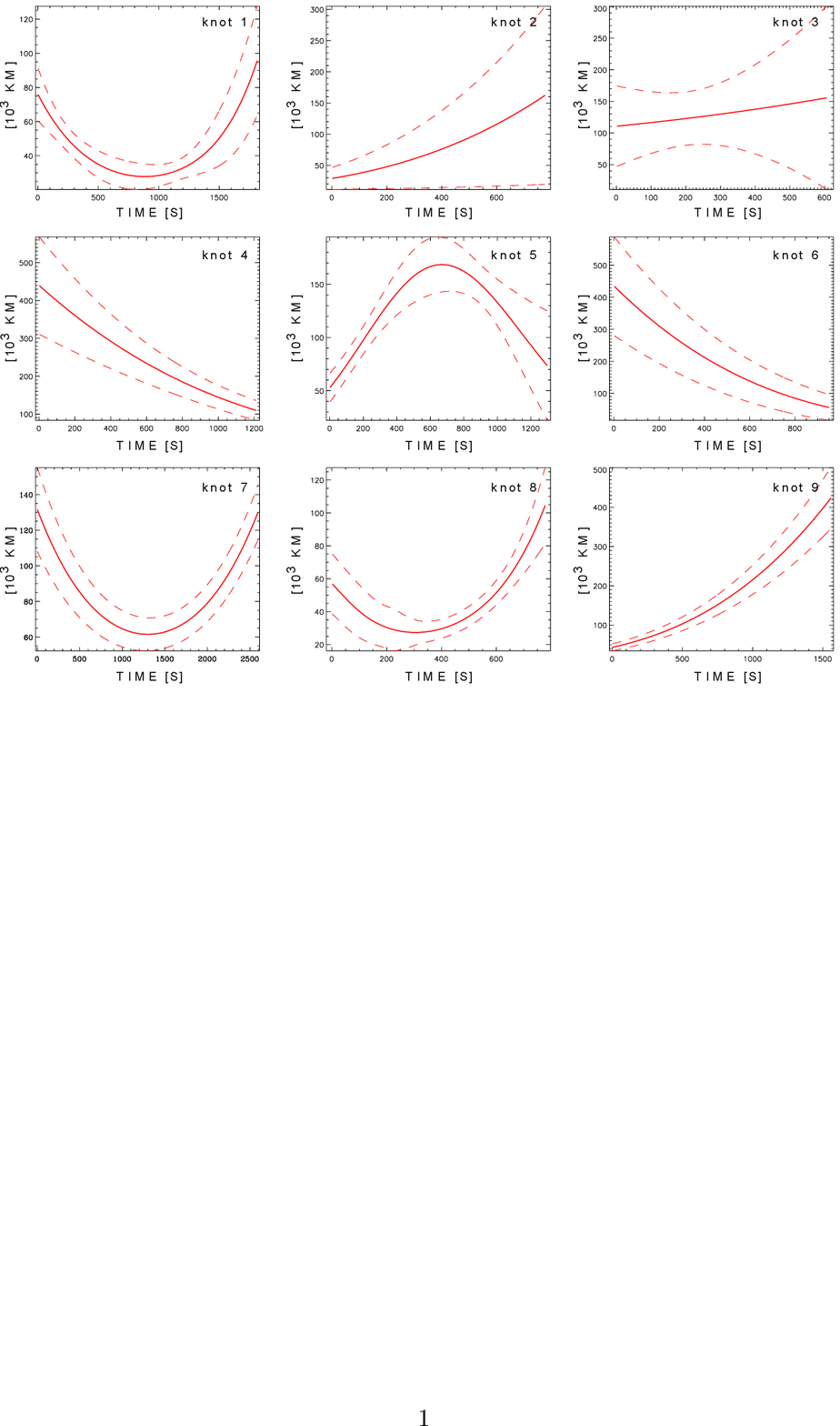}
\caption{Curvature radius (in units of $10^3$ km) of all observed knots. The horizontal axes  are scaled in seconds from the first time of observation of{ each} knot ($t_1$ - see Table \ref{TTT}). Dashed lines give error estimations from bootstrap resampling method. \label{CUR}}
\end{figure*}

\section{Discussions and conclusions}
\label{sec:conc}

During a coordinated campaign with space instruments IRIS and Hinode/SOT and the vector magnetograph THEMIS in the Canary Islands, a prominence{ was} observed on July 17 2014. This prominence is  a north-south oriented  filament but which is east-west extended, crossing the limb. The prominence{ appears} to have  a helical shape, with plasma turning along{ elliptical paths} { (as seen in Movies 1 and 2)} . 
We computed Doppler shifts  by taking the gravity centre of the IRIS \ion{Mg}{II} k line profiles, and  we obtained an anti-symmetric pattern with{ strong redshifts}
% and blueshifts
 around the axis of the prominence, suggesting a rotational motion around the axis,{ much like that seen in} AIA ``tornadoes'' \citep{Su14,Levens2015}.  { The spatial scale  has been degraded to 2\arcsec\ in order to have square pixels, and this explains why we do not see the Doppler shift of each individual knot.}

A 3D reconstruction{ technique}  \citep{Zapior2012} combining the IRIS slit jaw images and the high resolution spectra  of IRIS in{ the} \ion{Mg}{II} k line  allowed us{ to compute the velocity  vector  field in the volume of the prominence. The transverse velocity measured by using the SJI of IRIS is one order of magnitude less than the Doppler shifts. The apparent helical motion of the knots along the ellipses is very slow compared to the fast speed of the knots travelling along more or less horizontal threads with} 
%to understand that the prominence  consists  ofmore or less horizontal threads along which knots of plasma are travelling with 
velocities{ of} up to 65 km s$^{-1}$ away{ from the observer}.  The threads extended up to 80 000 km,  more or less  parallel{ to the solar surface,} and not along meridians in  planes without any torsion. { The apparent ellipse-shape structure could correspond to the curvature of the parallels of the Sun according to the value of the P angle in July (Figure \ref{Meudon} left top panel).}This extension corresponds to the { east-west extent}  of the filament a few days{ before} (around 5$^\circ$ in Figure \ref{Meudon}). The { dominant} force which  is{ moving the} knots is{ found to be} gas pressure  and  not gravity.

{ Using 3D reconstruction, we have demonstrated that  this    prominence is a typical prominence and does not have the characteristics of a tornado as we  thought it did at the beginning of the study. The high cadence of the movies  can give false impressions of the dynamics of these structures when projected onto the plane of the sky. We need to have simultaneous spectra and images to correctly interpret the movies.}

{ The horizontality of the threads forming the prominence  is consistent with the inclination of 90$^\circ$ found with THEMIS (Figure \ref{histo}).}
{ 
% It is perhaps abusing to compare results of different instruments. In coordinated observations we are commonly comparing data coming from different instruments with different spatial resolution and different cadence.  Putting together 
Combining data from two instruments 
%s always a source of trouble.
with different spatial resolution and different cadence is difficult and can lead to confusion when interpreting the observations.
However, THEMIS  is one of the few instruments able to measure 
%has the quasi-monopoly in measuring 
magnetic fields in prominences, together with the TIP instrument of the VTT. Such measurements come at a price: spatial and temporal resolutions are poor. Such information must always come from other instruments, SOT and IRIS  in the present work.
 It would  be certainly better  to measure magnetic fields while  maintaining an acceptable time cadence and a nice spatial resolution. A combination of adaptive optics, spectro-imagery and polarimetry would be desirable, if possible, in several spectral domains. 

We   have also to note  that the polarization is measured in  different lines  from  the  chromospheric lines in which prominences are observed.  
These lines have different characteristics (temperature, optical thickness) and they certainly formed in different structures. These  theoretical problems 
 should be resolved by non local thermodynamic equilibrium (NLTE)  modelling  of polarized light in the future.}

\begin{acknowledgements}

The authors thank S. Gunar, B. Gelly, and the team of THEMIS for assisting with the observations.  We would like to thank Jean-Marie Malherbe and Thierry Roudier who  { co-aligned the SOT frames. }They tested,   up to now unsuccessfully,   the CST and LCT  methods to  derive the flow pattern  in the plane of the sky.  We thank J. Dudik who provided the time-distance code to track the knots in the SOT cube of data, and  G.Aulanier and P.D\'emoulin for fruitful discussions. This paper was discussed during the ISSI workshop (PI N.Labrosse)  on "Solving the prominence paradox".
 P.J.L. acknowledges support from an STFC Research Studentship. N.L. acknowledges support from STFC grant ST/I001808/1. \textit{Hinode} is a Japanese mission developed and launched by ISAS/JAXA, with NAOJ as domestic partner and NASA and STFC (UK) as international partners. It is operated by these agencies in co-operation with ESA and NSC (Norway). IRIS is a NASA small explorer mission developed and operated by LMSAL with mission operations executed at NASA Ames Research Centre, and major contributions to downlink communications funded by the Norwegian Space Center (NSC, Norway) through an ESA PRODEX contract. The AIA data are provided courtesy of NASA/\textit{SDO} and the AIA science team.  M.Z. acknowledges support from the project RVO:67985815 of the Czech Academy of Sciences.

\end{acknowledgements}

%-------------------------------------------------------------------

%\bibliographystyle{aa}
%\bibliography{bibliography}

\end{document}